\documentclass{elsart}
\usepackage{psfrag,graphicx,float,array}
\usepackage{amsmath,amssymb}
\journal{hep-th}
\newcommand{\al}{\alpha}
\newcommand{\be}{\beta}
\newcommand{\de}{\delta}
\newcommand{\ep}{\epsilon}

\newcommand{\ga}{\gamma}
\newcommand{\ka}{\kappa}

\newcommand{\si}{\sigma}

\newcommand{\vt}{\vartheta}
\newcommand{\ze}{\zeta}
\newcommand{\De}{\Delta}

\newcommand{\La}{\Lambda}
\newcommand{\Sig}{\Sigma}
\newcommand{\Om}\Omega
\newcommand{\bk}{\mathbf{k}}
\newcommand{\bs}{\mathbf{s}}
\newcommand{\bx}{\mathbf{x}}
\newcommand{\bnu}{\boldsymbol{\nu}}

\newcommand{\bJ}{\mathbf{J}}
\newcommand{\bZ}[4]{\mathbf{Z}^{(#1|#2)}_{#3#4}}
\newcommand{\tK}{\tilde{K}}
\newcommand{\tS}{\tilde{S}}
\newcommand{\tc}{\tilde{c}}
\newcommand{\tm}{\tilde{m}}
\newcommand{\tn}{\tilde{n}}
\newcommand{\tilh}{\tilde{h}}

\def\CC{\mathbb{C}}
\def\NN{\mathbb{N}}
\def\RR{\mathbb{R}}

\newcommand{\cE}{{\mathcal E}}

\newcommand{\cH}{{\mathcal H}}

\newcommand{\cP}{{\mathcal P}}

\newcommand{\cU}{{\mathcal U}}
\newcommand{\cX}{{\mathcal X}}
\newcommand{\cY}{{\mathcal Y}}

\newcommand{\cZ}{{\mathcal Z}}
\def\ket#1{|#1\rangle}

\let\ni\noindent
\newcommand{\ms}{\mspace{1mu}}

\renewcommand{\le}{\leqslant}
\renewcommand{\ge}{\geqslant}
\renewcommand{\leq}{\leqslant}
\renewcommand{\geq}{\geqslant}
\newcommand{\qsb}[2]{[\ms #1\ms]_{#2}}
\newcommand{\qbinom}[3]{\genfrac{[}{]}{0pt}{}{\,#1\,}{#2}_{#3}}
\newcommand{\comp}{\mathbin{\raise.3ex\hbox{$\circ$}}}
\newcommand{\smax}{s_{\mathrm{max}}}
\newcommand{\smin}{s_{\mathrm{min}}}
\newcommand{\erf}{\operatorname{erf}}

\newcommand{\tr}{\operatorname{tr}}

\newcommand\Hsc{H_{\mathrm{sc}}}

\newcommand\Zsc{Z_{\mathrm{sc}}}

\newcommand\Sij{S_{ij}^{(m|n)}}
\newcommand\Sil{S_{il}^{(m|n)}}
\newcommand\Skl{S_{kl}^{(m|n)}}
\newcommand\tSij{\tilde S_{ij}^{(m|n)}}
\newcommand\tSil{\tilde S_{il}^{(m|n)}}
\newcommand\tSkl{\tilde S_{kl}^{(m|n)}}
\newcommand\Sji{S_{ji}^{(m|n)}}
\newcommand\Si{S_i^{\ep\ep'}}
\newcommand\Sj{S_j^{\ep\ep'}}
\newcommand\Sk{S_k^{\ep\ep'}}
\newcommand\Piij{\Pi_{ij}^{(m|n)}}
\newcommand\Pii{\Pi_i^{\ep\ep'}}
\newcommand\Pimn[2]{\Pi_{#1#2}^{(m|n)}}
\newcommand\Piee[1]{\Pi_{#1}^{\ep\ep'}}
\newcounter{ex}
\def\cond{\stepcounter{ex}\hskip-.75cm
\makebox[.55cm][r]{(\roman{ex})}\hskip.2cm}

\begin{document}
\begin{frontmatter}
\title{An exactly solvable supersymmetric spin chain of $BC_N$ type}
\author{J.C. Barba}, \author{F. Finkel}, \author{A.
  Gonz\'{a}lez-L\'{o}pez}, \author{M.A. Rodr\'{\i}guez}
\address{Depto.~de F\'{\i}sica Te\'{o}rica II, Universidad
Complutense, 28040 Madrid, Spain}
\date{July 9, 2008}

\begin{abstract}
We construct a new exactly solvable supersymmetric spin chain related to the
$BC_N$ extended root system, which includes as a particular case the
$BC_N$ version of the Polychronakos--Frahm spin chain. We also introduce
a supersymmetric spin dynamical model of Calogero type which yields
the new chain in the large coupling limit. This connection is exploited
to derive two different closed-form expressions for
the chain's partition function by means of Polychronakos's freezing trick.
We establish a boson-fermion duality relation for the new chain's spectrum,
which is in fact valid for a large class of (not necessarily integrable)
spin chains of $BC_N$ type. The exact expressions for the partition function are
also used to study the chain's spectrum as a whole, showing that
the level density is normally distributed even for a moderately large number of particles.
We also determine a simple analytic approximation to the distribution of normalized spacings
between consecutive levels which fits the numerical data with remarkable accuracy. Our results
provide further evidence that spin chains of Haldane--Shastry type are exceptional integrable models,
in the sense that their spacings distribution is not Poissonian, as posited by the Berry--Tabor
conjecture for ``generic'' quantum integrable systems.
\end{abstract}

\end{frontmatter}

\section{Introduction}\label{sec:Intro}

In the last few years exactly solvable (or integrable) supersymmetric spin chains and their associated
dynamical models have been the subject of extensive research in connection with different topics of
current interest, such as the theory of strongly correlated systems~\cite{ASK01,AS06} or the AdS-CFT
correspondence~\cite{HL04}. Among these models, the supersymmetric versions of the celebrated
Haldane--Shastry chain~\cite{Ha88,Sh88} and its rational counterpart proposed by Polychronakos~\cite{Po93}
and Frahm~\cite{Fr93} occupy a distinguished position, due to the rich mathematical structures at the heart of
their highly solvable character. The Haldane--Shastry (HS) chain was introduced in an attempt
to construct a simple one-dimensional model whose ground state coincided with Gutzwiller's variational wave function
for the Hubbard model in the limit of large on-site interaction~\cite{Hu63,Gu63,GV87}. It can also be obtained
in this limit from the Hubbard model with long-range hopping studied in Ref.~\cite{GR92}, in the half-filling
regime. The original HS chain describes $N$ spin $1/2$ particles equally spaced on a circle,
with pairwise interactions
inversely proportional to the square of the chord distance. Its Hamiltonian can be written as
\begin{equation}\label{HS}
\cH=J_0\sum_{i\ne j} \sin(\vt_i-\vt_j)^{-2}\bJ_i\cdot\bJ_j\,,\qquad\vt_i\equiv\frac{i\pi}N\,,
\end{equation}
where $\bJ_i\equiv\frac12\,(\si_i^x,\si_i^y,\si_i^z)$, $\si^\al_i$ is a Pauli matrix at site $i$,
and the summation indices run from $1$ to $N$ (as always hereafter, unless otherwise stated). A natural generalization
of this chain to su($m$) spin~\cite{Ka92,HH92} is obtained by taking $\bJ_i=(J_i^1,\dots,J_i^{m^2-1})$, where
$\{J_i^\al\}$ are the generators of the fundamental representation of su($m$) at site
$i$. In fact, with the usual normalization $\tr(J^\al_kJ^{\ga}_k)=\frac12\de^{\al\ga}$ we have
\[
\bJ_i\cdot\bJ_j=\frac12\,\Big(S_{ij}-\frac1m\Big)\,,
\]
where $S_{ij}$ is the operator permuting the $i$-th and $j$-th spins, so that the su($m$)
spin Hamiltonian~\eqref{HS} is a linear combination of spin permutation operators.

The spectrum of the spin $1/2$ HS Hamiltonian was
numerically analyzed in the original papers of Haldane and Shastry.
The more general su($m$) spin case was studied in a subsequent publication
by Haldane et al.~\cite{HHTBP92}, who empirically found a complete description of the spectrum
and explained its high degeneracy by an underlying $\cY(\mathrm{sl}_m)$ Yangian symmetry.
These results were rigorously established in
Ref.~\cite{BGHP93} by constructing a transfer matrix using
the Dunkl operators~\cite{Du89,Po92} of the (trigonometric)
Sutherland dynamical model~\cite{Su71,Su72}. Although this approach yields
an explicit formula for the energies in terms of the so-called {\em motifs},
the computation of their corresponding degeneracies becomes quite cumbersome when $m>2$.
This is probably the reason why the partition function of this chain was computed only
very recently~\cite{FG05}, using a procedure known as Polychronakos's {\em freezing trick} which
does not rely on the explicit knowledge of the spectrum~\cite{Po93,Po94}.

The key idea behind the freezing trick is exploiting the connection between the HS chain and the Sutherland spin
dynamical model in the strong coupling limit. Indeed, in this limit the particles in the latter model
``freeze'' at the coordinates of the (unique) equilibrium of the scalar part of the potential,
which are exactly the chain sites. Thus, in this limit the dynamical and the spin degrees of freedom decouple, so that the spectrum of the spin Sutherland model becomes approximately the sum of the spectra of the scalar Sutherland model and the HS chain. This observation yields an explicit formula for the partition function of the HS spin chain as the strong coupling limit of the quotient of the partition functions of the spin and the scalar Sutherland models,
both of which can be computed in this limit. In fact, this method was first applied by Polychronakos to evaluate the partition
function of the su($m$) spin chain associated with the spin Calogero (rational) model~\cite{Ca71,MP93}, usually
known in the literature as the Polychronakos--Frahm (PF) chain, whose Hamiltonian (in the antiferromagnetic case)
is given by
\begin{equation}\label{PF}
\cH=\sum_{i<j}(\ze_i-\ze_j)^{-2}(1+S_{ij})\,.
\end{equation}
The sites $\ze_i$ of this chain are the coordinates of the equilibrium of the scalar part of the Calogero potential,
which coincide with the zeros of the Hermite polynomial of degree $N$~\cite{Ca77}.

Both the HS and the PF chains admit a natural su($m|n$) supersymmetric version, in which there are
$m$ bosonic and $n$ fermionic degrees of freedom at each site~\cite{BUW99,HB00,BB06,BBHS07,BBS08}.
In practice, the Hamiltonians of these chains are obtained by replacing the spin permutation
operators $S_{ij}$ by suitable supersymmetric generalizations thereof (cf.~Eq.~\eqref{Sij} below).
The partition functions of the su($m|n$) PF and HS chains have also been computed in closed form
using the freezing trick method outlined above~\cite{BUW99,BB06}. These partition functions
were then used to establish a remarkable boson-fermion duality relation for the spectrum of each of these
chains~\cite{BUW99,BBHS07}.

The original HS and PF chains (and their supersymmetric versions) are associated with the $A_{N-1}$
root system, since the interaction between any two spins depends only on the difference of their site
coordinates. It is also of interest to consider generalizations of these chains
to the $BC_N$ extended root system, in which the spins interact not only among themselves but also with their
mirror images with respect to a reflecting end located at the origin. In the non-supersymmetric case,
the HS chain of $BC_N$ type was first discussed in Ref.~\cite{BPS95}, and its partition function was evaluated
by means of the freezing trick in Ref.~\cite{EFGR05}. Likewise, the partition function
of the su($m$) PF chain of $BC_N$ type introduced in~\cite{YT96} was recently computed in closed form
by the authors~\cite{BFGR08}. To the best of our knowledge, however, the supersymmetric
counterparts of both the HS and PF chains of $BC_N$ type have not been studied so far.
One of the main purposes of this paper is precisely to fill this gap in the case of the PF chain.
More precisely, we shall apply the freezing trick to compute the partition function of this chain
in closed form, relating it to the corresponding partition function of the su($m|n$)
PF chain of $A_{N-1}$ type. We shall also extend the boson-fermion duality mentioned above to a large class
of supersymmetric spin chains of $BC_N$ type which includes the PF chain as a particular case.

The spectra of the non-supersymmetric spin chains of Haldane--Shastry type whose
partition function has been explicitly computed
share some statistical properties that we shall now discuss. In the first place, the density of energy levels
follows the Gaussian law with remarkable accuracy even for a moderately large number of
particles~\cite{FG05,EFGR05,BFGR08,BFGR08b}. Secondly
(except for the HS chain of $BC_N$ type, which has not yet been studied in this respect), the density
of spacings $p(s)$ between consecutive (normalized) levels follows a characteristic distribution
essentially different from Poisson's law $p(s)=\e^{-s}$. This fact is somewhat surprising, in view of
the integrability of these chains~\cite{Po93,YT96} and the long-standing conjecture of Berry and Tabor~\cite{BT77},
according to which the spacings distribution of a ``generic'' quantum integrable system should be Poissonian.
The validity of this conjecture has been verified for a number of important integrable
systems, including the Heisenberg chain, the $t\ms$-$J$ model, the
Hubbard model~\cite{PZBMM93}, and the chiral Potts model~\cite{AMV02}.
On the other hand, the non-Poissonian character of the spacings distribution has recently been observed
for the su($m|n$) HS chain of $A_{N-1}$ type~\cite{BB06}, whose density of spacings
is qualitatively similar to that of the non-supersymmetric HS and PF chains. Another important
aim of this paper is the study of the level density and the spacings distribution for
the su($m|n$) PF chain of $BC_N$ type. Our main conclusion in this respect is that this chain behaves in essentially
the same way as other chains of Haldane--Shastry type previously studied. In other words,
when the number of spins is sufficiently large the level density follows with great accuracy
the Gaussian distribution, and the spacings distribution also deviates substantially from Poisson's law.
As a matter of fact, we shall see that the spacings density is well approximated by the same ``square root of a logarithm'' distribution as the original HS and PF chains, and the su($m$) PF chain of $BC_N$ type~\cite{BFGR08,BFGR08b}.
This lends further support to the fact that spin chains of Haldane--Shastry type are exceptional
integrable systems from the point of view of the Berry--Tabor conjecture.

The paper is organized as follows. In Section~\ref{sec:themodel} we recall the definition
of the supersymmetric spin permutation operators~\cite{BUW99}, and introduce new
supersymmetric spin reversal operators. With the help of these operators, we construct the su($m|n$)
PF chain of $BC_N$ type and explain how the freezing trick can be used to compute its
partition function from that of its associated spin dynamical model. In Section~\ref{sec:spectrum}
we compute the spectrum of the latter model, which is then used in Section~\ref{sec:partition}
to evaluate the chain's partition function. Using the grand canonical partition function of the spin
dynamical model, in Section~\ref{sec:rel}
we obtain an alternative expression for the chain's partition function in terms of that of
the su($m|n$) PF chain of $A_{N-1}$ type. This expression turns out to be particularly efficient
for the computation of the chain's spectrum for a relatively large number of spins. Section~\ref{sec:dual}
is devoted to generalizing to spin chains
of $BC_N$ type the boson-fermion duality uncovered in Refs.~\cite{BUW99,BBHS07}.
In the last section we use the explicit formulas for the partition function obtained
in Section~\ref{sec:rel} to study the asymptotic behavior of the level density and the spacings distribution
when the number of particles is very large. The paper also includes two appendices, the first one being
a collection of several useful $q$-number identities, and the second a detailed derivation of the
exact formulas for the mean and standard deviation of the chain's energy used in Section~\ref{sec:lev}.

\section{The model}\label{sec:themodel}

Let us begin by defining the Hilbert space of the internal degrees of freedom
and the action therein of the $\mathrm{su}(m|n)$ permutation and spin reversal operators
$\Sij$ and $\Si$. There are two species of particles, ``bosons'' and
``fermions'' (the reason for this terminology shall be apparent later), with
$m$ and $n$ denoting respectively the number of bosonic and fermionic degrees
of freedom. More formally, \emph{the internal Hilbert space}
$\Sigma^{(m|n)}\approx(\CC^{m+n})^{\otimes N}$ is spanned by states of the
form
$\ket{s_1}\otimes\cdots\otimes\ket{s_N}\equiv\ket{s_1,\dots,s_N}\equiv\ket\bs\,,$
where $s_i$ is a two-vector $(s_i^1,s_i^2)$ with

\ni\hphantom{i}i) $s_i^2\equiv\pi(s_i)=
\begin{cases}
  0,\quad & \text{bosons}\\ 1,& \text{fermions}
\end{cases}$

\smallskip
\ni ii) $s_i^1\in
\begin{cases}
  \big\{-\frac{m-1}2,-\frac{m-1}2+1,\dots,\frac{m-1}2\big\}\,,\quad & \pi(s_i)=0\\[1mm]
   \big\{-\frac{n-1}2,-\frac{n-1}2+1,\dots,\frac{n-1}2\big\}\,,\quad & \pi(s_i)=1\,.
\end{cases}
$

\ni We shall say that $s_i^1$ is the \emph{value} of the
``spin'' $s_i$, while $s_i^2\equiv\pi(s_i)$ is its \emph{type}
(bosonic or fermionic). As usual, the inner product on $\Sig^{(m|n)}$ is defined
in such a way that the set of vectors $\ket\bs$ forms an orthonormal basis.

The action of the \emph{spin reversal operators} $\Si$ on this basis is given by
\begin{equation}\label{Si}
\Si\ket{s_1,\dots,s_i,\dots,s_N}=\rho(s_i)\,\ket{s_1,\dots,s_i^-,\dots,s_N}\,,
\end{equation}
where $s_i^-=(-s_i^1,s_i^2)$,
\[
\rho(s_i)=\begin{cases}
	\ep\,, &\pi(s_i)=0\\
	\ep'\,, &\pi(s_i)=1\,,
\end{cases}
\]
and $\ep,\ep'=\pm1$ are two independent signs. In other words, $\Si$ reverses the value of
the $i$-th spin without affecting its type, and multiplies the state by the sign $\rho(s_i)$.

The definition of the \emph{spin permutation operators} $\Sij\equiv\Sji$ is
a bit more involved, namely (assuming that $i<j$)
\begin{equation}\label{Sij}
\Sij\ket{\dots,s_i,\dots,s_j,\dots} =
(-1)^{\alpha_{ij}(\bs)}\ket{\dots,s_j,\dots,s_i,\dots}\,,
\end{equation}
where
\[
\alpha_{ij}(\bs)=\pi(s_i)\pi(s_j)+\big(\pi(s_i)+\pi(s_j)\big)f_{ij}(\bs)\,,
\]
and
\[
f_{ij}(\bs)=\sum_{k=i+1}^{j-1}\pi(s_k)
\]
is the number of fermions in between the $i$-th and $j$-th spins.
In other words, the sign $(-1)^{\alpha_{ij}(\bs)}$ is $-1$ when either both
$s_i$ and $s_j$ are fermionic, or $s_i$ and $s_j$ are spins of different type
with an odd number of fermionic spins between them. An equivalent way
of defining the action of $\Sij$ is by requiring that $S_{i,i+1}^{(m|n)}$
introduce a factor of $-1$ precisely when $s_i$ and $s_{i+1}$ are fermionic.
This clearly implies the previous (apparently more general) rule, since an
arbitrary permutation is a product of permutations of consecutive elements.

The Hamiltonian of the supersymmetric su($m|n$) PF chain of $BC_N$ type is
defined by\footnote{Note that the operators $\tS_{ij}^{(m|n)}$ actually depend on $\ep$ and $\ep'$,
although for simplicity we have chosen not to make this dependence explicit in the notation.}
\begin{equation}\label{cH}
\cH^{(m|n)}_{\ep\ep'}=\sum_{i\neq j}\bigg[
\frac{1-S_{ij}^{(m|n)}}{(\xi_i-\xi_j)^2}
+\frac{1-\tS_{ij}^{(m|n)}}{(\xi_i+\xi_j)^2}\bigg]
+\be\sum_i\frac{1-S_i^{\ep\ep'}}{\xi_i^2}\,,
\end{equation}
where $\tS_{ij}^{(m|n)}=S_i^{\ep\ep'}S_j^{\ep\ep'}S_{ij}^{(m|n)}$ and $\be>0$.
The chain sites $\xi_i$ are the same as those of the ordinary PF chain of $BC_N$ type
studied in Ref.~\cite{BFGR08}, namely the coordinates of the (unique) equilibrium of
the scalar potential
\begin{equation}\label{U}
U(\bx)=\sum_{i\neq j}\bigg[\frac1{(x_i-x_j)^2}+\frac1{(x_i+x_j)^2}\bigg]
+\sum_i\frac{\be^2}{x_i^2}+\frac{r^2}4
\end{equation}
in the open set $C=\{\bx\equiv(x_1,\dots,x_N)\mid 0<x_1<\dots<x_N\}$.
It can be shown~\cite{YT96} that $\xi_i=\sqrt{2y_i}$, where $y_i$ is the $i$-th zero of
the Laguerre polynomial $L_N^{\be-1}$.
Note also that the convention for the sign of the SUSY spin
permutation operators in the definition of $\cH^{(m|n)}_{\ep\ep'}$ is the
opposite to that of Ref.~\cite{BB06}, but (as we shall see in a moment)
our choice is more consistent with the use of the names ``bosonic'' and
``fermionic'' for the two species of spins. Since
\begin{equation}\label{Sij0}
S_{ij}^{(m|0)}=S_{ij}\,,\qquad S_{ij}^{(0|n)}=-S_{ij}\,,\qquad S_i^{\ep\ep}=\ep S_i\,,
\end{equation}
where $S_i$ is the operator reversing the $i$-th spin, the purely bosonic (resp.~purely fermionic) spin
chain $\cH^{(m|0)}_{\ep}$
(resp.~$\cH^{(0|n)}_{\ep}$) coincides with the ordinary
ferromagnetic (resp.~antiferromagnetic) PF chain of $BC_N$ type~\cite{BFGR08}.

The spin chain~\eqref{cH} is naturally related to the su($m|n$) spin dynamical model
\begin{equation}\label{H}
H^{(m|n)}_{\ep\ep'}=-\triangle+a\sum_{i\neq j}\bigg[
\frac{a-\Sij}{(x_{ij}^-)^2}+\frac{a-\tS_{ij}^{(m|n)}}{(x_{ij}^+)^2}\bigg]
+b\sum_i\frac{b-\Si}{x_i^2}+\frac{a^2}4\,r^2\,,
\end{equation}
where $x_{ij}^\pm=x_i\pm x_j$, $b=\be a$ and $a>1/2$, and to its scalar counterpart
\begin{equation}\label{Hsc}
H_{\text{sc}}=-\triangle +a(a-1)\sum_{i\neq j}\bigg[
\frac1{(x_{ij}^-)^2}+\frac1{(x_{ij}^+)^2}\bigg]
+b(b-1)\sum_i\frac1{x_i^2}+\frac{a^2}4\,r^2\,.
\end{equation}
Indeed, we have
\begin{equation}\label{HHH}
H^{(m|n)}_{\ep\ep'}=H_{\text{sc}}+a\,\tilde\cH^{(m|n)}_{\ep\ep'}\,,
\end{equation}
where $\tilde\cH^{(m|n)}_{\ep\ep'}$ is obtained from the spin chain Hamiltonian~\eqref{cH}
by replacing the sites $\xi_i$ by the coordinates $x_i$. Since
\[
H^{(m|n)}_{\ep\ep'}=-\triangle+a^2 U+O(a)\,,
\]
as $a$ grows to infinity the particles tend to concentrate at the chain sites $\xi_i$,
so that the dynamical and internal degrees of freedom decouple. By Eq.~\eqref{HHH},
in this limit the energies of the dynamical spin model~\eqref{H} are approximately given by
\begin{equation}\label{EEE}
E_{ij}\simeq E^{\mathrm{sc}}_i+a\cE_j\,,\qquad a\gg1\,,
\end{equation}
where $E^{\mathrm{sc}}_i$ and $\cE_j$ are any two energies of the scalar Hamiltonian~\eqref{Hsc}
and the spin chain~\eqref{cH}. The latter formula cannot be used directly to deduce the
spectrum of $\cH^{(m|n)}_{\ep\ep'}$ from those of $H^{(m|n)}_{\ep\ep'}$ and $H_{\text{sc}}$,
since it is not known {\em a priori} which eigenvalues of $H^{(m|n)}_{\ep\ep'}$ and $H_{\text{sc}}$ combine to
yield an eigenvalue of $\cH^{(m|n)}_{\ep\ep'}$. However, Eq.~\eqref{EEE} immediately
leads to the exact ``freezing trick'' formula~\cite{Po94}
\begin{equation}\label{ZZZ}
\cZ^{(m|n)}_{\ep\ep'}(T)=\lim_{a\to\infty}\frac{Z^{(m|n)}_{\ep\ep'}(aT)}{Z_{\mathrm{sc}}(aT)}
\end{equation}
relating the partition functions $\cZ^{(m|n)}_{\ep\ep'}$, $Z^{(m|n)}_{\ep\ep'}$ and
$Z_{\text{sc}}$ of $\cH^{(m|n)}_{\ep\ep'}$, $H^{(m|n)}_{\ep\ep'}$ and
$H_{\text{sc}}$, respectively. The partition function $Z_{\text{sc}}$ of the scalar model~\eqref{Hsc}
was computed in Ref.~\cite{BFGR08}. We shall see in the next section that
the spectrum of $H^{(m|n)}_{\ep\ep'}$ is quite simple, a fact that shall be exploited in
Section~\ref{sec:partition} to explicitly compute its partition
function. Using Eq.~\eqref{ZZZ} we shall then obtain
the partition function $\cZ^{(m|n)}_{\ep\ep'}$ in closed form.

\section{Spectrum of the spin dynamical model}\label{sec:spectrum}

The first step in the evaluation of the partition function of
the spin chain~\eqref{cH} by means of the freezing trick relation~\eqref{ZZZ}
consists in determining the spectrum of the spin dynamical model~\eqref{H}.
To this end, we introduce the auxiliary ``exchange Hamiltonian''
\begin{equation}\label{Hp}
H'=-\triangle+\sum_{i\neq j}\bigg[
\frac a{(x_{ij}^-)^2}(a-K_{ij})+\frac a{(x_{ij}^+)^2}(a-\tK_{ij})\bigg]
+\sum_i\frac b{x_i^2}\,(b-K_i)+\frac{a^2}4\,r^2\,,
\end{equation}
where $K_{ij}$ is the operator permuting the $i$-th and $j$-th spatial coordinates,
$K_i$ reverses the sign of the $i$-th coordinate,
and $\tK_{ij}\equiv K_iK_jK_{ij}$. The rationale for introducing this
operator is the fact that its restriction to states symmetric under
the action of the operators
\begin{equation}
\Piij=K_{ij}\Sij\,,\qquad \Pii=K_i\Si\,,\qquad i,j=1,\dots,N\,,
\end{equation}
coincides with that of the Hamiltonian $H^{(m|n)}_{\ep\ep'}$, as we shall next discuss. 
To begin with, note that all three sets of operators $\{K_i,K_{ij}\}$, $\{\Si,\Sij\}$ and
$\{\Pii,\Piij\}$ generate a realization of the Weyl group of
$BC_N$ type, namely they satisfy the commutation relations
\begin{equation}
\begin{gathered}
    K_{ij}^2=1,\qquad K_{ij}K_{jk}=K_{ik}K_{ij}=K_{jk}K_{ik},\qquad
    K_{ij}K_{kl}=K_{kl}K_{ij},\\[1mm]
    K_i^2=1,\qquad K_iK_j=K_jK_i\,,\qquad K_{ij}K_k=K_k K_{ij},\qquad
    K_{ij}K_j=K_i K_{ij}\,,
\end{gathered}
    \label{BCN}
\end{equation}
(where $i,j,k,l$ are distinct) and similarly for the other two sets. We shall denote by $\La^{(m|n)}_{\ep\ep'}$
the projector on states totally symmetric under the action of the operators
$\{\Pii,\Piij\}$. For instance, if $N=3$ we have
\begin{align*}
\La^{(m|n)}_{\ep\ep'}&=\frac1{2^3}(1+\Piee1)(1+\Piee2)(1+\Piee3)\\
&\times\frac1{3!}\big(1+\Pimn12+\Pimn13+\Pimn23+\Pimn12\Pimn13+\Pimn12\Pimn23\big)\,.
\end{align*}
In particular, $\La^{(m|0)}_{\ep\ep}$ (resp.~$\La^{(0|n)}_{\ep\ep}$) projects
onto states totally symmetric (resp. antisymmetric) under particle permutations and
with parity $\ep$ with respect to simultaneous sign reversals of spatial coordinates and spins.
{}From the definition of the total symmetrizer $\La^{(m|n)}_{\ep\ep'}$ it follows that
\[
\Piij\La^{(m|n)}_{\ep\ep'}=\La^{(m|n)}_{\ep\ep'}\Piij=\La^{(m|n)}_{\ep\ep'}\,,\qquad
\Pii\La^{(m|n)}_{\ep\ep'}=\La^{(m|n)}_{\ep\ep'}\Pii=\La^{(m|n)}_{\ep\ep'}\,,
\]
so that
\begin{equation}\label{SK}
\Sij\La^{(m|n)}_{\ep\ep'}=K_{ij}\La^{(m|n)}_{\ep\ep'}\,,\qquad
\Si\La^{(m|n)}_{\ep\ep'}=K_i\La^{(m|n)}_{\ep\ep'}\,.
\end{equation}
Hence
\begin{equation}\label{HHpLambda}
H^{(m|n)}_{\ep\ep'}\La^{(m|n)}_{\ep\ep'}= H'\La^{(m|n)}_{\ep\ep'}\,,
\end{equation}
which, as we shall see, is precisely the relation needed to deduce the spectrum of
$H^{(m|n)}_{\ep\ep'}$ from that of $H'$.

Indeed, recall~\cite{BFGR08} that the operator
$H'$ is represented by an upper triangular matrix in the (non-orthonormal)
basis with elements
\begin{equation}
\phi_\bk=\e^{-\frac a4\ms r^2}\prod_i|x_i|^b\cdot\prod_{i<j}{\big|x_i^2-x_j^2\big|}^a\cdot\prod_ix_i^{k_i}\,,
\end{equation}
where
\[
\bk\equiv(k_1,\dots,k_N)\in{(\NN\cup\{0\})}^N\equiv\NN_0^N\,.
\]
In other words,
\[
H'\phi_\bk=E'_\bk\phi_\bk+\sum_{\vert\bk'\vert<\vert\bk\vert}c_{\bk'\bk}\phi_{\bk'}\,,
\]
where the eigenvalues are given by
\[
E'_\bk=a\ms\vert\bk\vert+E_0\,,
\]
with $\vert\bk\vert=\sum_i k_i$ and $E_0=Na\big(b+a(N-1)+\frac12\big)$.

Due to the impenetrable nature of the singularities of the Hamiltonian~\eqref{H},
its Hilbert space is the set $L_0^2(C)\otimes\Sig^{(m|n)}$ of spin functions square integrable on the open set $C$
and vanishing sufficiently fast on the singular hyperplanes $x_i\pm x_j=0$. However, it can be shown that
$H^{(m|n)}_{\ep\ep'}$ is equivalent to its symmetric extension (under $\Piij$ and $\Pii$)
to the space $L_0^2(\RR^N)\otimes\Sig^{(m|n)}$. The states
\begin{equation}\label{psis}
\psi_{\bk,\bs}=\La^{(m|n)}_{\ep\ep'}\big(\phi_\bk(\bx)\ket\bs\big)\,,
\end{equation}
form a basis of this Hilbert space provided that the following conditions hold:

\begin{minipage}{\textwidth}
{\leavevmode\hphantom{ii}(i)\enspace $k_1\geq\cdots\geq k_N$\\[1.5mm]
\leavevmode\hphantom{i}(ii)\enspace If $k_i=k_j$ and $i<j$, then:\\[1mm]
\leavevmode\hphantom{(iii)}\enspace (ii.a)\enspace $\pi(s_i)\leq\pi(s_j)$\\[1mm]
\leavevmode\hphantom{(iii)}\enspace (ii.b)\enspace $s_i^1\geq s^1_j+\pi(s_j)$
whenever $\pi(s_i)=\pi(s_j)$\\[1.5mm]
(iii)\enspace $s_i^1\geq 0$ for all $i$, and $s_i^1>0$ if $(-1)^{k_i}\rho(s_i)=-1$}
\end{minipage}

The last two conditions, which differ from the analogous
ones for the $\mathrm{su}(m)$ case~\cite{BFGR08}, deserve further remark.
Condition (ii.a) simply states that acting with
appropriate generalized spin permutation operators $\Sij$
we can first reorder the spins so that
within each \emph{sector} of $\bs$ (spin components
corresponding to a maximal sequence of equal components of $\bk$)
bosons precede fermions. Furthermore (condition (ii.b)),
the bosonic spin components can be arranged in
non-increasing order, while for the fermionic ones we can enforce strictly
decreasing order since in this case the state vanishes if $s_i=s_j$ by
antisymmetry. (This is, by the way, the justification for the names
``bosonic'' and ``fermionic'' used for the two species of spins.)
Finally, condition (iii) stems from the fact that the spin functions~\eqref{psis}
must be even under $\Pii$, while if $s_i^1=0$ we have
\[
\Pii\psi_{\bk,\bs}=\La^{(m|n)}_{\ep\ep'}\big(\Pii\phi_\bk(\bx)\ket\bs\big)
=(-1)^{k_i}\rho(s_i)\,\psi_{\bk,\bs}\,,
\]
so that $(-1)^{k_i}\rho(s_i)=1$. A direct consequence of this condition
is that the number of
bosonic spin values of $s_j$ compatible with rule (iii) is given by
\begin{equation}\label{mbar}
\overline m(k_j)=
\begin{cases}
  \frac m2\,,& m \text{ even}\\[1mm]
  \frac{m+\ep}2\,,\text{\quad}& m \text{ odd, $k_j$ even}\\[1mm]
  \frac{m-\ep}2\,,& m \text{ odd, $k_j$ odd\,.}
\end{cases}
\end{equation}
Similarly, the number $\overline n(k_j)$ of fermionic spin values of $s_j$ compatible with the
third rule is given by the previous expression, with $m$ replaced by $n$
and $\ep$ by~$\ep'$.

The Hamiltonian $H^{(m|n)}_{\ep\ep'}$ is represented by an upper triangular
matrix in the basis~\eqref{psis}, ordered according to the degree $\vert\bk\vert$. Indeed,
taking into account that $H'$ commutes with $\La^{(m|n)}_{\ep\ep'}$, by Eq.~\eqref{HHpLambda} we have
\[
H^{(m|n)}_{\ep\ep'}\psi_{\bk,\bs}=H'\psi_{\bk,\bs}
=\La^{(m|n)}_{\ep\ep'}\big((H'\phi_\bk)\ket\bs\big)
=E'_\bk\psi_{\bk,\bs}+\sum_{\vert\bk'\vert<\vert\bk\vert}c_{\bk'\bk}\psi_{\bk',\bs}\,.
\]
Hence the eigenvalues of $H^{(m|n)}_{\ep\ep'}$ are given by
\begin{equation}\label{Eks}
E_{\bk,\bs}=a\vert\bk\vert+E_0\,,
\end{equation}
where $\bk$ and $\bs$ satisfy conditions (i)--(iii) above. Note that, although
the value of $E_{\bk,\bs}$ is independent of $\bs$, the degeneracy of each
level clearly depends on the spin through the latter conditions.

{}From Eqs.~\eqref{Sij0}--\eqref{Hsc} it follows that the
Hamiltonian~$\Hsc$ of the scalar Calogero model is simply $H^{(1|0)}_{+}$,
so that its spectrum is also given by the RHS of Eq.~\eqref{Eks}, with all $k_i$ even by
condition (iii).
We can thus subtract the ground energy $E_0$ from the spectra of both
$\Hsc$ and $H^{(m|n)}_{\ep\ep'}$ without altering
the partition function of the chain~\eqref{cH}. With this proviso,
the energies of both models are proportional to the coupling constant~$a$,
which implies that $\Zsc(aT)$ and $Z^{(m|n)}_{\ep\ep'}(aT)$ are independent of $a$.
Hence, in the calculation of the partition function of the chain~\eqref{cH}
we can take without loss of generality $E_0=0$ and $a=1$, and write the
freezing trick formula~\eqref{ZZZ} simply as
\begin{equation}\label{ZZZsimp}
\cZ^{(m|n)}_{\ep\ep'}(T)=\frac{Z^{(m|n)}_{\ep\ep'}(T)}{Z_{\mathrm{sc}}(T)}\,.
\end{equation}

\section{Evaluation of the partition function}\label{sec:partition}

We are now ready to compute the partition function of the spin chain~\eqref{cH} using the freezing trick
formula~\eqref{ZZZsimp}. As for the ordinary (purely bosonic or fermionic) chain, the evaluation of the partition
function $Z^{(m|n)}_{\ep\ep'}$ of the spin dynamical model~\eqref{H}
depends on the parity of the integers $m$ and $n$,
so that one has to consider four different cases. In the calculations that follow,
we shall use the fact that the number of ways into which we can arrange the spin
components $s_i$ in a sector of $\bs$ corresponding to a certain component
$\ka$ of $\bk$ repeated $\nu$ times is given by
\begin{equation}\label{dnu}
d(\nu,\ka)= \sum_{i=0}^{\min(\overline n(\ka),\nu)}
\binom{\overline n(\ka)}{i}\binom{\overline m(\ka)+\nu-i-1}{\nu-i}\,.
\end{equation}
Clearly, this number depends on $\ka$ only through its parity, cf.~\eqref{mbar}
and its fermionic analogue. Note also that when $\overline m(\ka)=0$
the second binomial coefficient in~\eqref{dnu} should be interpreted
as $\de_{i\nu}$.

\subsection*{Case 1: $m,n$ even}

Recall, to begin with, that after setting $E_0=0$ and $a=1$
the partition function of the scalar model~\eqref{Hsc} reads~\cite{BFGR08}
\begin{equation}\label{Zsc}
\Zsc(T)=\sum_{k_1\geq\cdots\geq k_N\geq 0}q^{2\vert\bk\vert}
=\prod_i(1-q^{2i})^{-1}\,,\qquad q\equiv\e^{-1/(k_{\mathrm B}T)}\,.
\end{equation}
By Eq.~\eqref{Eks}, the partition function of the Hamiltonian~\eqref{H} can be written as
\begin{equation}\label{Z}
Z^{(m|n)}_{\ep\ep'}(T)=\sum_{k_1\geq\cdots\geq k_N\geq 0}d_\bk\ms q^{\vert\bk\vert}\,,
\end{equation}
where the \emph{spin degeneracy factor} $d_\bk$ is the number of spin states
$\ket\bs$ satisfying conditions (ii) and (iii) for a given multi-index $\bk$.
Since both $m$ and $n$ are even, by Eq.~\eqref{mbar} and its fermionic analogue we have
$\overline m(k_j)=m/2$ and $\overline n(k_j)=n/2$. If
\begin{equation}\label{bk}
\bk=(\overbrace{\vphantom{1}\ka_1,\dots,\ka_1}^{\nu_1},\dots,
\overbrace{\vphantom{1}\ka_r,\dots,\ka_r}^{\nu_r}),\qquad\ka_1>\cdots>\ka_r\geq0,
\end{equation}
from Eq.~\eqref{dnu} it follows that
\begin{equation}\label{dbk}
d_\bk=\prod\limits_{j=1}^r d(\nu_j)\equiv d(\bnu)\,,
\end{equation}
where $\bnu=(\nu_1,\dots,\nu_r)$ and
\begin{equation}
\label{deven}
d(\nu_j)=\sum_{i=0}^{\min(\frac n2,\,\nu_j)}\binom{\frac n2}{i}\binom{\frac m2+\nu_j-i-1}{\nu_j-i}\,.
\end{equation}
Since $\sum_{i=1}^r\nu_i=N$, the multi-index $\bnu$ is an element of the set $\cP_N$
of partitions of $N$ with order taken into account. Inserting Eqs.~\eqref{bk}--\eqref{deven}
into~\eqref{Z} we obtain
\[
Z^{(m|n)}_{\ep\ep'}=\sum_{\bnu\in\cP_N}d(\bnu)\sum_{\ka_1>\cdots>\ka_r\geq 0}\,q^{\sum\limits_{i=1}^r\nu_i\ka_i}\,.
\]
The RHS of the previous equation can be evaluated as in~\cite[Eq.~(24)]{BFGR08}, with the result
\begin{equation}\label{Zee}
Z^{(m|n)}_{\ep\ep'}=q^{-N}\sum_{\bnu\in\cP_N}d(\bnu)\prod_{j=1}^{\ell(\bnu)}\frac{q^{N_j}}{1-q^{N_j}}\,,
\end{equation}
where $\ell(\bnu)=r$ is the number of components of the multi-index $\bnu$ and
\[
N_j=\sum_{i=1}^j\nu_i\,.
\]
{}From Eqs.~\eqref{ZZZsimp}, \eqref{Zsc} and~\eqref{Zee} it follows that in this case the partition function
of the $\mathrm{su}(m|n)$ PF chain of $BC_N$ type is given by
\begin{equation}\label{cZee}
\cZ^{(m|n)}_{\ep\ep'}=\prod_i(1+q^{i})\cdot\sum_{\bnu\in\cP_N}d(\bnu)\,q^{\sum\limits_{j=1}^{\ell(\bnu)-1}\kern-5pt
N_j}\,\prod_{j=1}^{N-\ell(\bnu)}\big(1-q^{N_j'}\big)\,,\qquad m,n\in 2\NN_0\,,
\end{equation}
where the positive integers $N_j'$ are defined by
\[
\big\{N_1',\dots,N'_{N-\ell(\bnu)}\big\} = \big\{1,\dots,N-1\big\}
-\big\{N_1,\dots,N_{\ell(\bnu)-1}\big\}\,.
\]
Note that, as for the purely bosonic/fermionic model, in this case the
partition function does not depend on $\ep,\ep'$. We shall therefore drop from
now on the subscripts $\ep,\ep'$ from $\cZ^{(m|n)}_{\ep\ep'}$ when both $m$ and
$n$ are even. The first product in Eq.~\eqref{cZee} is simply the partition function
of the su$(2|0)$ chain~\eqref{cH} (cf.~Eq.~(34) in Ref.~\cite{BFGR08}), whereas
the remaining sum coincides with the partition function $\cZ^{(\frac m2|\frac n2)}_{(\text A)}$
of the su$(\frac m2|\frac n2)$ PF chain of type $A_{N-1}$ computed in~\cite{BBS08}. We thus obtain
the remarkable relation
\begin{equation}
\label{releven}
\cZ^{(m|n)} = \cZ^{(2|0)}\,\cZ^{(\frac m2|\frac n2)}_{(\text A)}\,,\qquad m,n\in 2\NN_0\,.
\end{equation}

Recall~\cite{BUW99} that the partition function of the su$(k|l)$ PF chain of type A is explicitly
given by
\begin{equation}\label{cZAkl}
\cZ^{(k\ms|l)}_{\mathrm{(A)}}=\sum_{M_1+\cdots+M_{k+l}=N}q^{\frac12\sum\limits_{j=k+1}^{k+l}M_j(M_j-1)}
\frac{(q)_N}{(q)_{M_1}\cdots(q)_{M_{k+l}}}\,,
\end{equation}
where the notation $(q)_k$ is defined in Eq.~\eqref{qkkq} of Appendix~\ref{sec:q}.
For instance, from Eq.~\eqref{cZAkl} with $k=l=1$ and Eqs.~\eqref{qkkq}--\eqref{qbinom} one obtains
\[
\cZ^{(1|1)}_{(\text A)}=\sum_{M=0}^N \qbinom NMq\,q^{\binom M2}=\prod_{i=0}^{N-1}(1+q^i)\,,
\]
so that
\begin{equation}\label{cZ22}
\cZ^{(2|2)}=\cZ^{(2|0)}\,\cZ^{(1|1)}_{(\text A)}=2(1+q^N)\prod_{i=1}^{N-1}(1+q^i)^2\,.
\end{equation}

\subsection*{Case 2: $m$ odd, $n$ even}

In this case, by Eq.~\eqref{mbar} and its fermionic counterpart we have $\overline n(k_j)=n/2$,
while $\overline m(k_j)$ depends on the parity of $k_j$. Hence it is convenient
to modify condition (i) by first grouping separately the components of $\bk$ with the same
parity and then ordering separately the even and odd components. In other words, we shall now
write  $\bk=(\bk_{\mathrm e},\bk_{\mathrm o})$, where
\begin{align*}
&\bk_{\mathrm e}=\big(\overbrace{\vphantom{1}2\ka_1,\dots,2\ka_1}^{\nu_1},\dots,
\overbrace{\vphantom{1}2\ka_s,\dots,2\ka_s}^{\nu_s}\big),\\
&\bk_{\mathrm o}=\big(\overbrace{\vphantom{1}2\ka_{s+1}+1,\dots,2\ka_{s+1}+1}^{\nu_{s+1}},\dots,
\overbrace{\vphantom{1}2\ka_r+1,\dots,2\ka_r+1}^{\nu_r}\big),
\end{align*}
and
\[
\ka_1>\cdots>\ka_s\geq0,\qquad\ka_{s+1}>\cdots>\ka_r\geq0\,.
\]
{}By Eq.~\eqref{mbar}, $\overline m(2\ka_j)=(m+\ep)/2$ for $j=1,\dots,s$, while $\overline m(2\ka_j+1)=(m-\ep)/2$
for $j=s+1,\dots,r$. {}From Eq.~\eqref{dnu} it follows that the spin degeneracy factor $d_\bk$ is now
given by
\begin{equation}\label{depsoe}
d_\bk=\prod_{j=1}^sd_\ep(\nu_j)\cdot
\prod_{j=s+1}^r d_{-\ep}(\nu_j)\equiv d^s_\ep(\bnu)\,,
\end{equation}
where
\begin{equation}\label{doe}
d_\ep(\nu_j)=\sum_{i=0}^{\min(\frac n2,\,\nu_j)}
\binom{\frac n2}{i}\binom{\frac{m+\ep}2+\nu_j-i-1}{\nu_j-i}\,.
\end{equation}
Equation~\eqref{Z} for the partition function of the dynamical spin model~\eqref{H} then becomes
\[
Z^{(m|n)}_{\ep\ep'}=\sum_{\bnu\in\cP_N}\sum_{s=0}^r d_\ep^s(\bnu)
\hspace*{-10pt}\sum_{\substack{\ka_1>\cdots>\ka_s\geq 0\\ \ka_{s+1}>\cdots>\ka_r\geq0}}\hspace*{-10pt}
q^{\sum\limits_{i=1}^s2\nu_i\ka_i\;+\!\sum\limits_{i=s+1}^r\nu_i(2\ka_i+1)}\,,
\]
where the RHS can be evaluated as in~\cite[Eq.~(29)]{BFGR08}. We thus obtain
\begin{equation}\label{Zoe}
Z^{(m|n)}_{\ep\ep'}=\sum_{\bnu\in\cP_N}\sum_{s=0}^{\ell(\bnu)}d_\ep^s(\bnu)\,q^{-(N+N_s)}
\prod_{j=1}^s\frac{q^{2N_j}}{1-q^{2N_j}}
\cdot\prod_{j=s+1}^{\ell(\bnu)}\frac{q^{2\tilde
N_j}}{1-q^{2\tilde N_j}}\,,
\end{equation}
where
\[
\tilde N_j = \sum_{i=s+1}^j\nu_{i}\,,\qquad j=s+1,\dots,\ell(\bnu)\,.
\]
Substituting Eqs.~\eqref{Zsc} and~\eqref{Zoe} into~\eqref{ZZZsimp} we arrive at the following explicit
expression for the partition function of the spin chain~\eqref{cH} for odd $m$ and even~$n$:
\begin{equation} \label{cZoe}
\cZ^{(m|n)}_{\ep\ep'}=\prod_i(1-q^{2i})\cdot\sum_{\bnu\in\cP_N}
\sum_{s=0}^{\ell(\bnu)}d_\ep^s(\bnu)\,q^{-(N+N_s)}
\prod_{j=1}^s\frac{q^{2N_j}}{1-q^{2N_j}}\,\,\cdot\!\!
\prod_{j=s+1}^{\ell(\bnu)}\frac{q^{2\tilde N_j}}{1-q^{2\tilde N_j}}\,.
\end{equation}
Note that in this case the partition function depends on $\ep$ but not on $\ep'$.

As is the case for any finite system with integer energies, the
partition function $\cZ^{(m|n)}_{\ep\ep'}$ should be a polynomial in $q$.
We shall now prove this fact by simplifying Eq.~\eqref{cZoe} with the help of the identities
in Appendix~\ref{sec:q}. To this end, we define the two sets of integers
\begin{align*}
&\big\{N_1',\dots,N'_{N_s-s}\big\} = \big\{1,\dots,N_s-1\big\}
-\big\{N_1,\dots,N_{s-1}\big\}\,,\\
&\big\{\tilde N_{N_s-s+1}',\dots,\tilde N'_{N-\ell(\bnu)}\big\} = \big\{1,\dots,N-N_s-1\big\}
-\big\{\tilde N_{s+1},\dots,\tilde N_{\ell(\bnu)-1}\big\}\,,
\end{align*}
in terms of which we can rewrite Eq.~\eqref{cZoe} as
\begin{multline} \label{cZoesimp}
\cZ^{(m|n)}_{\ep\ep'}=\sum_{\bnu\in\cP_N}
\sum_{s=0}^{\ell(\bnu)}d_\ep^s(\bnu)\,q^{N-N_s+2\sum\limits_{j=1}^{s-1}N_j
+2\sum\limits_{j=s+1}^{\ell(\bnu)-1}\tilde N_j}\,\qbinom N{N_s}{q^2}\\
\times\prod_{j=1}^{N_s-s}(1-q^{2N'_j})\,\cdot\!
\prod_{j=N_s-s+1}^{N-\ell(\bnu)}(1-q^{2{\tilde N}'_j})\,.
\end{multline}
{}From the discussion in Appendix~\ref{sec:q} (cf.~Eqs.~\eqref{qbinompoly}--\eqref{tau}), it
follows that the binomial coefficient $\qbinom N{N_s}{q^2}$ is an even polynomial of
degree $2N_s(N-N_s)$ in $q$. Thus the RHS of Eq.~\eqref{cZoesimp} is also a polynomial in $q$,
as expected.

\subsection*{Case 3: $m$ even, $n$ odd}

The evaluation of the partition function $\cZ^{(m|n)}_{\ep\ep'}$ is performed as in
the previous case, the only difference being that the spin degeneracy factor $d_\ep^s(\bnu)$
in Eq.~\eqref{cZoesimp} should be replaced by
\begin{equation}\label{depseo}
d^s_{\ep'}(\bnu)=\prod_{j=1}^sd_{\ep'}(\nu_j)\cdot
\prod_{j=s+1}^r d_{-\ep'}(\nu_j)\,,
\end{equation}
where now
\begin{equation}\label{deo}
d_{\ep'}(\nu_j)=\sum_{i=0}^{\min(\frac{n+\ep'}2,\,\nu_j)}
\binom{\frac {n+\ep'}2}{i}\binom{\frac m2+\nu_j-i-1}{\nu_j-i}\,.
\end{equation}
In particular, the partition function $\cZ^{(m|n)}_{\ep\ep'}$ does not depend on $\ep$ in this case.

\subsection*{Case 4: $m,n$ odd}

In this case both $\overline m(k_j)$ and $\overline n(k_j)$ depend on the parity of $k_j$,
so that the computation of the partition function proceeds as in Case~2.
The final result is still Eq.~\eqref{cZoesimp}, with the spin degeneracy factor $d_\ep^s(\bnu)$
replaced by
\begin{equation}\label{depsoo}
d^s_{\ep\ep'}(\bnu)=\prod_{j=1}^sd_{\ep\ep'}(\nu_j)\cdot
\prod_{j=s+1}^r d_{-\ep,-\ep'}(\nu_j)\,,
\end{equation}
where
\begin{equation}\label{doo}
d_{\ep\ep'}(\nu_j)=\sum_{i=0}^{\min(\frac {n+\ep'}2,\,\nu_j)}
\binom{\frac {n+\ep'}2}{i}\binom{\frac{m+\ep}2+\nu_j-i-1}{\nu_j-i}\,.
\end{equation}
Thus in this case the partition function depends on both $\ep$ and $\ep'$.

\section{Relation with the spin chain of $A_{N-1}$ type}\label{sec:rel}

We have seen in the previous section that when the integers $m$ and $n$ are both even,
the partition function $\cZ^{(m|n)}$ of the chain~\eqref{cH} is related in a simple way
to that of the $\mathrm{su}(\frac m2|\frac n2)$ PF chain of $A_{N-1}$ type, cf.~Eq.~\eqref{releven}.
The purpose of this section is to derive an analogous relation for odd values of $m$ and/or $n$.
To this end, we shall deduce an alternative expression for the partition function
by means of the grand canonical partition function of the spin dynamical model~\eqref{H},
following an approach similar to that of Refs.~\cite{Po94,BUW99} for the ordinary and supersymmetric
PF chains of $A_{N-1}$ type.

In order to compute the grand canonical partition function, it is convenient to
replace conditions (i)--(iii) in Section~\ref{sec:spectrum} ordering the labels $\bk,\bs$
of the basis states~\eqref{psis} by the following equivalent set of rules:

\begin{tabbing}
(a)\enspace \= If $i<j$ then $\pi(s_i)\le\pi(s_j)$\\[1mm]
(b)\enspace \> If $i<j$ and $\pi(s_i)=\pi(s_j)$, then $s_i^1\le s_j^1$\\[1mm]
(c)\enspace \> If $i<j$ and $s_i=s_j$, then $k_i\ge k_j+\pi(s_j)$\\[1mm]
(d)\enspace \> $s_i\ge0$, and $(-1)^{k_i}=\rho(s_i)$ if $s_i^1=0$.
\end{tabbing}

By Eq.~\eqref{Eks} (setting, as before, $E_0=0$ and $a=1$) the grand canonical partition function
$\bZ mn\ep{\ep'}$ of the spin dynamical model~\eqref{H} is given by
\begin{equation}\label{bZ}
\bZ mn\ep{\ep'}(T,\mu)=\sum_{N=0}^\infty\sum_{\bk,\bs}q^{\vert\bk\vert-N\mu}\,,
\end{equation}
where $\mu$ is the chemical potential and the inner sum runs over all values of $\bk\in\NN_0^N$ and $\bs$
compatible with conditions (a)--(d).
As in the previous section, the computation of $\bZ mn\ep{\ep'}$
depends on the parity of $m$ and $n$. Since the case of even $m$ and $n$ has already been dealt with
above, we shall start with the case of $m$ odd and $n$ even:

\subsection*{Case 2: $m$ odd, $n$ even}

Calling $\tm=(m-1)/2$ and $\tn=n/2$, by conditions (a), (b), and (d) we can write
\begin{equation}\label{bs}
\bs=\big(\overbrace{\vphantom{\tfrac{1}2}0,\dots,0}^{\nu_0}\,,\dots,
\overbrace{\vphantom{\tfrac{1}2}\tfrac{m-1}2,\dots,\tfrac{m-1}2}^{\nu_{\tm}}\,,
\overbrace{\tfrac{1}2,\dots,\tfrac{1}2}^{\nu_{\tm+1}}\,,\dots,
\overbrace{\vphantom{\tfrac{1}2}\tfrac{n-1}2,\dots,\tfrac{n-1}2}^{\nu_{\tm+\tn}}\big)\,,
\end{equation}
where  $\nu_i\in\NN_0$ and $\sum_{i=0}^{\tm+\tn}\nu_i=N$.
With a slight abuse of notation, we have suppressed the second component $s_i^2$
of each spin $s_i$, since clearly $s_i^2=0$ for the first $\nu_0+\dots+\nu_{\tm}$ spins
and $s_i^2=1$ for the remaining ones. By condition~(c), the multi-index $\bk$ is of the form
\begin{equation}\label{bkoe}
\bk=\big(2\ka^0_1+p\,,\dots,2\ka^0_{\nu_0}+p\,,\ka^1_1,\dots,\ka^1_{\nu_1},
\dots,\ka^{\tm+\tn}_1,\dots,\ka^{\tm+\tn}_{\nu_{\tm+\tn}}\big)\,,
\end{equation}
where
\begin{equation}\label{p}
p=\frac12\,(1-\ep)
\end{equation}
and the nonnegative integers $\ka^i_j$ satisfy
\[
\begin{cases}
\ka^i_1\ge\cdots\ge\ka^i_{\nu_i}\,,&\quad i=0,\dots,\tm\\[1mm]
\ka^i_1>\cdots>\ka^i_{\nu_i}\,, &\quad i=\tm+1,\dots,\tm+\tn\,.
\end{cases}
\]
By Eq.~\eqref{bZ}, the grand canonical partition function of the chain~\eqref{cH}
is given by
\begin{align*}
\bZ mn\ep{\ep'}&=\sum_{\nu_0,\dots,\nu_{\tm+\tn}\ge0}\:
\sum_{\substack{\ka^i_1\ge\cdots\ge\ka^i_{\nu_i}\\i=0,\dots,\tm}}\,
\sum_{\substack{\ka^{j}_1>\cdots>\ka^{j}_{\nu_{j}}\\j=\tm+1,\dots,\tm+\tn}}
q^{\sum\limits_{l=1}^{\nu_0}(2\ka^0_l+p)-\nu_0 \mu}\,
\prod_{l=1}^{\tm+\tn}q^{{\sum\limits_{r=1}^{\nu_l}\ka^l_r-\nu_l \mu}}\\
&=\Bigg[\sum_{\substack{\ka^{0}_1\ge\cdots\ge\ka^{0}_{\nu_{0}}\\\nu_0\ge0}}
q^{\sum\limits_{l=1}^{\nu_0}(2\ka^0_l+p)-\nu_0 \mu}\Bigg]\,
\Bigg[\prod_{i=1}^{\tm}\sum_{\substack{\ka^{i}_1\ge\cdots\ge\ka^{i}_{\nu_{i}}\\\nu_i\ge0}}
q^{{\sum\limits_{r=1}^{\nu_i}\ka^i_r-\nu_i \mu}}
\Bigg]\\
&\hspace{14em}\times\Bigg[
\prod_{j=\tm+1}^{\tm+\tn}\sum_{\substack{\ka^{j}_1>\cdots>\ka^j_{\nu_j}\\\nu_j\ge0}}
q^{\sum\limits_{r=1}^{\nu_j}\ka^j_r-\nu_j\mu}\Bigg]\,.
\end{align*}
The first factor in the last equality is clearly the grand canonical partition function
$\bZ 10\ep{}$ of the purely bosonic spinless dynamical model~\eqref{H} (which coincides
with the scalar model~\eqref{Hsc} for $\ep=1$). On the other hand, in Ref.~\cite{BUW99}
it was shown that
\[
\sum_{\nu=0}^\infty\sum_{\ka_1\ge\cdots\ge\ka_\nu}
q^{{\sum\limits_{r=1}^\nu\ka_r-\nu\mu}}=\bZ 10{\mathrm{(A)}}{},\qquad
\sum_{\nu=0}^\infty\sum_{\ka_1>\cdots>\ka_\nu}
q^{{\sum\limits_{r=1}^\nu\ka_r-\nu\mu}}=\bZ 01{\mathrm{(A)}}{},
\]
where $\bZ 10{\mathrm{(A)}}{}$ (resp.~$\bZ 01{\mathrm{(A)}}{}$) denotes the grand canonical
partition function of the purely bosonic (resp.~fermionic) spinless Calogero model of type $A$.
We thus have
\[
\bZ mn\ep{\ep'}=\bZ 10\ep{}\ms\big(\bZ 10{\mathrm{(A)}}{}\big)^{\!\frac{m-1}2}
\big(\bZ 01{\mathrm{(A)}}{}\big)^{\!\frac n2}\,.
\]
Using again the results in Ref.~\cite{BUW99}, we can write the previous formula as
\begin{equation}\label{bZs}
\bZ mn\ep{\ep'}=\bZ 10\ep{}\,\bZ {\frac{m-1}2\ms}{\frac n2}{\mathrm{(A)}}{}\,,
\end{equation}
where $\bZ {\frac{m-1}2\ms}{\frac n2}{\mathrm{(A)}}{}$ denotes the grand canonical partition function
of the su$({\frac{m-1}2\ms}|{\frac n2})$ spin Calogero model of type $A$. On the other hand,
the grand canonical partition function $\mathbf Z(T,\mu)$ can be expressed in terms of the $N$-particle canonical
partition function $Z_N(T)$ as
\[
\mathbf Z(T,\mu)=\sum_{N=0}^\infty y^N Z_N(T)\,,
\]
where $y\equiv q^{-\mu}$ and $Z_0(T)\equiv1$. Equating the coefficient of
$y^N$ in both sides of Eq.~\eqref{bZs} we readily obtain
\[
Z^{(m|n)}_{\ep\ep',N}=\sum_{M=0}^N Z^{(1|0)}_{\ep,M}Z^{(\frac{m-1}2\ms|\frac n2)}_{\mathrm{(A)},N-M}\,,
\]
where $Z^{(\frac{m-1}2\ms|\frac n2)}_{\mathrm{(A)},N-M}$ is the partition function
of the su$({\frac{m-1}2\ms}|{\frac n2})$ spin Calogero model of $A_{N-1}$ type with $N-M$ particles.
By Eq.~\eqref{ZZZsimp}, the partition function of the $N$-particle su$(m|n)$
PF chain of $BC_N$ type is thus given by
\begin{equation}\label{cZN}
\cZ^{(m|n)}_{\ep\ep',N}=\sum_{M=0}^N \frac{Z^{(1|0)}_{\ep,M}}{Z_{\mathrm{sc},N}}
\,Z^{(\frac{m-1}2\ms|\frac n2)}_{\mathrm{(A)},N-M}\,.
\end{equation}
The partition function $Z^{(1|0)}_{+,M}$ is simply given by
\begin{equation}\label{Z10+M}
Z^{(1|0)}_{+,M}=Z_{\mathrm{sc},M}=\frac1{(q^2)_M}\,,
\end{equation}
where we have used~\eqref{Zsc}.
On the other hand, from conditions (i) and (iii) in Section~\ref{sec:spectrum} we have
\begin{equation}\label{Z10-M}
Z^{(1|0)}_{-,M}=\sum_{k_1\geq\cdots\geq k_M\geq 0}q^{\sum\limits_{i=1}^M (2k_i+1)}
=q^M Z_{\mathrm{sc},M}
=\frac{q^M}{(q^2)_M}\,.
\end{equation}
The partition function of the su$({\frac{m-1}2\ms}|{\frac n2})$ spin chain of type $A$ with
$N-M$ particles is related to the partition function of its corresponding spin dynamical model by~\cite{BUW99}
\begin{equation}\label{cZA}
\cZ^{(\frac{m-1}2\ms|\frac n2)}_{\mathrm{(A)},N-M}=(q)_{N-M}Z^{(\frac{m-1}2\ms|\frac n2)}_{\mathrm{(A)},N-M}\,.
\end{equation}
Inserting Eqs.~\eqref{Z10+M}--\eqref{cZA} into~\eqref{cZN} we finally obtain
\begin{align}
\cZ^{(m|n)}_{\ep\ep',N}&=\sum_{M=0}^N q^{\frac12(1-\ep)M}\frac{(q^2)_N}{(q^2)_M(q)_{N-M}}\,
\cZ^{(\frac{m-1}2\ms|\frac n2)}_{\mathrm{(A)},N-M}\notag\\
&=\sum_{M=0}^N q^{\frac12(1-\ep)M}\prod_{i=M+1}^N(1+q^i)\cdot\qbinom NMq\,
\cZ^{(\frac{m-1}2\ms|\frac n2)}_{\mathrm{(A)},N-M}\,,\label{cZoeA}
\end{align}
where we have used the identity~\eqref{qbinomid}. Alternatively, from the first equality in~\eqref{cZoeA}
and Eq.~\eqref{releven} we can also derive the relation
\[
\cZ^{(m|n)}_{\ep\ep',N}=\sum_{M=0}^N q^{\frac12(1-\ep)M}\qbinom NM{q^2}\,
\cZ^{(m-1|n)}_{N-M}\,.
\]

\subsection*{Case 3: $m$ even, $n$ odd}

This case is very similar to the previous one, the main difference being that
Eq.~\eqref{bZs} now becomes
\begin{equation}\label{bZseo}
\bZ mn\ep{\ep'}=\bZ 01{\ep'}{}\,\bZ {\frac{m}2\ms}{\frac{n-1}2}{\mathrm{(A)}}{}\,,
\end{equation}
which leads to the following expression for the chain's partition function:
\begin{equation}\label{cZNeo}
\cZ^{(m|n)}_{\ep\ep',N}=\sum_{M=0}^N \frac{Z^{(0|1)}_{\ep',M}}{Z_{\mathrm{sc},N}}
\,Z^{(\frac{m}2\ms|\frac{n-1}2)}_{\mathrm{(A)},N-M}\,.
\end{equation}
By conditions (i)--(iii) in Section~\ref{sec:spectrum}, the partition function $Z^{(0|1)}_{+,M}$ of
the spinless fermionic model~\eqref{H} with even parity is given by
\begin{multline}\label{Z01+M}
Z^{(0|1)}_{+,M}=\sum_{k_1>\cdots>k_M\geq 0}q^{2\sum\limits_{i=1}^M k_i}
=\sum_{l_1\ge\cdots\ge l_M\geq 0}q^{2\sum\limits_{i=1}^M (l_i+M-i)}\\
=q^{M(M-1)}Z_{\mathrm{sc},M}=\frac{q^{M(M-1)}}{(q^2)_M}\,.
\end{multline}
Similarly,
\begin{equation}\label{Z01-M}
Z^{(0|1)}_{-,M}=\sum_{k_1>\cdots>k_M\geq 0}q^{\sum\limits_{i=1}^M (2k_i+1)}
=q^M Z^{(0|1)}_{+,M}=\frac{q^{M^2}}{(q^2)_M}\,.
\end{equation}
Inserting~\eqref{Z01+M} and~\eqref{Z01-M} into Eq.~\eqref{cZNeo} and proceeding as before,
we obtain the following two expressions for the partition function of the chain~\eqref{cH}
with even $m$ and odd $n$:
\begin{subequations}
\begin{align}
\cZ^{(m|n)}_{\ep\ep',N}
&=\sum_{M=0}^N q^{M\left(M-\frac12(1+\ep')\right)}\prod_{i=M+1}^N(1+q^i)\cdot\qbinom NMq\,
\cZ^{(\frac{m}2\ms|\frac{n-1}2)}_{\mathrm{(A)},N-M}\label{cZeoA}\\[1mm]
&=\sum_{M=0}^N q^{M\left(M-\frac12(1+\ep')\right)}\qbinom NM{q^2}\,\cZ^{(m|n-1)}_{N-M}\label{cZeoB}\,.
\end{align}
\end{subequations}

\subsection*{Case 4: $m,n$ odd}

In this case both the bosonic and fermionic spin components can take the zero value, so that
instead of Eq.~\eqref{bZs} we now have
\begin{equation}\label{bZsoo}
\bZ mn\ep{\ep'}=\bZ 10{\ep}{}\,\bZ 01{\ep'}{}\,\bZ {\frac{m-1}2\ms}{\frac{n-1}2}{\mathrm{(A)}}{}\,.
\end{equation}
In particular, setting $m=n=1$ in the previous this formula we have
\begin{equation}\label{bZ11}
\bZ 11\ep{\ep'}=\bZ 10{\ep}{}\,\bZ 01{\ep'}{}\,,
\end{equation}
and therefore
\begin{equation}\label{bZsoo2}
\bZ mn\ep{\ep'}=\bZ 11{\ep}{\ep'}\,\bZ {\frac{m-1}2\ms}{\frac{n-1}2}{\mathrm{(A)}}{}\,.
\end{equation}
Proceeding again as in Case 2, we easily obtain
\begin{equation}\label{cZNoo}
\cZ^{(m|n)}_{\ep\ep',N}=\sum_{M=0}^N \frac{Z^{(1|1)}_{\ep\ep',M}}{Z_{\mathrm{sc},N}}
\,Z^{(\frac{m-1}2\ms|\frac{n-1}2)}_{\mathrm{(A)},N-M}
=\sum_{M=0}^N\frac{(q^2)_N}{(q^2)_M(q)_{N-M}}
\,\cZ^{(1|1)}_{\ep\ep',M}\,\cZ^{(\frac{m-1}2\ms|\frac{n-1}2)}_{\mathrm{(A)},N-M}\,.
\end{equation}
As before, this expression can be simplified in two alternative ways, leading to the following
remarkable identities for the partition function of the chain~\eqref{cH} with odd $m$ and $n$:
\begin{subequations}\label{cZooAB}
\begin{align}
\cZ^{(m|n)}_{\ep\ep',N}
&=\sum_{M=0}^N\,\prod_{i=M+1}^N(1+q^i)\cdot\qbinom NMq\,\cZ^{(1|1)}_{\ep\ep',M}\,
\cZ^{(\frac{m-1}2\ms|\frac{n-1}2)}_{\mathrm{(A)},N-M}\label{cZooA}\\[1mm]
&=\sum_{M=0}^N \qbinom NM{q^2}\,\cZ^{(1|1)}_{\ep\ep',M}\,\cZ^{(m-1|n-1)}_{N-M}\label{cZooB}\,.
\end{align}
\end{subequations}

Let us finally evaluate the partition function $\cZ^{(1|1)}_{\ep\ep'}$ appearing in Eqs.~\eqref{cZooAB}.
To begin with, from Eq.~\eqref{bZ11} for the grand canonical partition function and the freezing trick
formula~\eqref{ZZZsimp} we obtain
\begin{equation}\label{cZ11}
\cZ^{(1|1)}_{\ep\ep'}=\sum_{M=0}^N\frac{Z^{(1|0)}_{\ep,N-M}Z^{(0|1)}_{\ep',M}}{Z_{\mathrm{sc},N}}
=\sum_{M=0}^N q^{p(N-M)+M(M-1+p')}\qbinom NM{q^2}\,,
\end{equation}
where $p$ is given by~\eqref{p} and $p'=(1-\ep')/2$. We thus have
\begin{equation}\label{cZ11++}
\cZ^{(1|1)}_{++}=\sum_{M=0}^N q^{2\binom M2}\qbinom NM{q^2}=2\prod_{i=1}^{N-1}(1+q^{2i})\,,
\end{equation}
where we have used Eq.~\eqref{qbinom} with $x=1$ and $q$ replaced by $q^2$. On the other hand,
if $\ep=\ep'=-1$ Eq.~\eqref{cZ11} becomes
\begin{equation}\label{cZ11--}
\cZ^{(1|1)}_{--}=q^N\sum_{M=0}^N q^{2\binom M2}\qbinom NM{q^2}=q^N\cZ^{(1|1)}_{++}\,.
\end{equation}
Next, if $\ep=-\ep'=1$ we have
\begin{equation}\label{cZ11+-}
\cZ^{(1|1)}_{+-}=\sum_{M=0}^N q^{M^2}\qbinom NM{q^2}=\prod_{i=0}^{N-1}(1+q^{2i+1})\,,
\end{equation}
where we have used Eq.~\eqref{qbinom2}. Finally, when $\ep=-\ep'=-1$, Eqs.~\eqref{cZ11} and~\eqref{qbinom3}
yield
\begin{multline}\label{cZ11-+}
\cZ^{(1|1)}_{-+}=q^N\sum_{M=0}^N q^{M(M-2)}\qbinom NM{q^2}=q^N\prod_{i=0}^{N-1}(1+q^{2i-1})\\
=q^{N-1}(1+q)^2\prod_{i=2}^{N-1}(1+q^{2i-1})\,.
\end{multline}

\section{Boson-fermion duality}\label{sec:dual}

In Ref.~\cite{BBHS07} it was uncovered a remarkable boson-fermion duality
satisfied by the spectrum of any spin chain of $A_{N-1}$ type with global su$(m|n)$ symmetry.
We shall now extend these ideas to the PF chain of $BC_N$ type~\eqref{cH}, although our
results will in fact be valid for any chain of $BC_{N}$ type possessing global su$(m|n)$ symmetry.

Let us start by defining the \emph{star operator}
${}^*:\Sig^{(m|n)}\to\Sig^{(m|n)}$ by
\begin{equation}\label{star}
\ket\bs^* = (-1)^{\sum_ii\ms\pi(s_i)}\ms\ket\bs\,,
\end{equation}
and the \emph{exchange operator} $\cX^{(m|n)}:\Sig^{(m|n)}\to\Sig^{(n|m)}$ by
\begin{equation} \label{exc}
\cX^{(m|n)}\ket{s_1,\dots,s_N}=\ket{s_1',\dots,s_N'}\,,\qquad s'_i\equiv(s_i^1,1-s_i^2)\,.
\end{equation}
In other words, $s_i'$ has the same value as $s_i$ but opposite type.
Following Ref.~\cite{BBHS07}, we next define the operator $\cU^{(m|n)}:\Sig^{(m|n)}\to\Sig^{(n|m)}$
as the composition
\begin{equation}\label{cU}
\cU^{(m|n)}=\cX^{(m|n)}\comp{}^*\,.
\end{equation}
Since, clearly, the operators ${}^*\comp{}^*$ and $\cX^{(n|m)}\comp\cX^{(m|n)}$ are the
identity in $\Sig^{(m|n)}$, $\cU^{(m|n)}$ is invertible and
\[
{\cU^{(m|n)}}^{-1}={}^*\comp\cX^{(n|m)} = (-1)^{\frac12N(N+1)}\,\cU^{(n|m)}.
\]
Indeed, if $\ket\bs\in\Sig^{(n|m)}$, we have
\begin{multline*}
\ket{s_1',\dots,s_N'}^*=(-1)^{\sum_ii\ms\pi(s_i')}\ket{s_1',\dots,s_N'}=
(-1)^{\sum_ii(1-\pi(s_i))}\ket{s_1',\dots,s_N'}\\=
(-1)^{\sum_ii}\,\cU^{(n|m)}\ket\bs=(-1)^{\frac12N(N+1)}\,\cU^{(n|m)}\ket\bs\,.
\end{multline*}
Moreover, it is also easy to see that the star operator
is self-adjoint and ${\cX^{(m|n)}}^\dagger=\cX^{(n|m)}$, so that
$
{\cU^{(m|n)}}^\dagger = {}^*\comp\cX^{(n|m)}={\cU^{(m|n)}}^{-1}\,,
$
i.e., $\cU^{(m|n)}$ is unitary. A straightforward computation~\cite{BBHS07} shows that $\cU^{(m|n)}
S_{ij}^{(m|n)} = -S_{ij}^{(n|m)}\cU^{(m|n)}$, or equivalently
\begin{equation}\label{SijcU}
{\cU^{(m|n)}}^{-1}S_{ij}^{(n|m)}\,\cU^{(m|n)}=-S_{ij}^{(m|n)}\,.
\end{equation}
The above considerations are enough for proving duality in the $A_N$ case.
However, in the $BC_N$ case we need also consider the
behavior of the spin reversal operators $S_i^{\ep\ep'}$ with respect to conjugation by
$\cU^{(m|n)}$. For clarity's sake, in the following discussion we shall use the more precise
notation $S^{(m\ep|n\ep')}_i$ to denote the operator $S_i^{\ep\ep'}:\Sig^{(m|n)}\to\Sig^{(m|n)}$\,.
Noting that $(s_i^-)'=(s_i')^-$ and $\pi(s_i^-)=\pi(s_i)$, if $\ket\bs\in\Sig^{(m|n)}$ we have
\begin{align*}
\cU^{(m|n)}\,&S_i^{(m\ep|n\ep')}\,\ket{\bs}=\rho(s_i)\,\cU^{(m|n)}\ket{s_1,\dots,s_i^-,\dots,s_N}\\
&\quad=\rho(s_i)\,(-1)^{\sum_ii\,\pi(s_i)}\,\ket{s'_1,\dots,(s_i^-)',\dots,s'_N}\\
&\quad=\rho(s_i)\,(-1)^{\sum_ii\,\pi(s_i)}\,\ket{s'_1,\dots,(s'_i)^-,\dots,s'_N}
\equiv S_i^{(n\ep'|m\ep)}\,\cU^{(m|n)}\,\ket\bs\,,
\end{align*}
and therefore
\begin{equation}\label{SicU}
{\cU^{(m|n)}}^{-1}S_i^{(n\ep'|m\ep)}\ms\cU^{(m|n)}=S_i^{(m\ep|n\ep')}=-S_i^{(m,-\ep\mid n,-\ep')}\,.
\end{equation}
With the help of Eqs.~\eqref{SijcU} and~\eqref{SicU} one easily obtains
\begin{multline*}
\cH^{(m|n)}_{-\ep,-\ep'}+{\cU^{(m|n)}}^{-1}\cH^{(n|m)}_{\ep'\ep}\,\cU^{(m|n)}
\\=2\Big(\sum_{i\neq j}\big[(\xi_i-\xi_j)^{-2}+(\xi_i+\xi_j)^{-2}\big]
+\be\sum_i\xi_i^{-2}\Big)=N^2,
\end{multline*}
where the last sum was evaluated in Ref.~\cite{BFGR08}.
Since ${\cU^{(m|n)}}^{-1}\cH^{(n|m)}_{\ep'\ep}\,\cU^{(m|n)}$ and $\cH^{(n|m)}_{\ep'\ep}$ are
isospectral, from the previous equation we obtain the remarkable duality relation
\begin{equation}\label{dual}
\cZ^{(n|m)}_{\ep'\ep}(q)=q^{N^2}\cZ^{(m|n)}_{-\ep,-\ep'}(q^{-1})\,.
\end{equation}
When $\ep'=\ep$, the previous formula relates the spectra of two chains which differ in the exchange
of bosons with fermions and the action of the spin reversal operators. On the other hand,
when $\ep'=-\ep$ Eq.~\eqref{dual} is a genuine duality relation with respect to the exchange of
bosons and fermions, i.e., it relates the spectra of the chains $\cH^{(m|n)}_{-\ep,\ep}$ and $\cH^{(n|m)}_{-\ep,\ep}$.
In addition, if $m=n$ Eq.~\eqref{dual} establishes that the spectrum of the chain $\cH^{(m|m)}_{-\ep,\ep}$
is invariant under the transformation $\cE\mapsto N^2-\cE$, i.e., it
is symmetric about $N^2/2$. For instance, for $m=1$ this property can be
checked directly from Eqs.~\eqref{cZ11+-}-\eqref{cZ11-+} for the partition function.

Similarly, if $m$ and $n$ are both even the partition function does not depend on $\ep$ and $\ep'$, so that
Eq.~\eqref{dual} is again a genuine boson-fermion duality relation. For the same reason, if $m=n\in2\NN_0$
the spectrum of the chain $\cH^{(m|m)}_{\ep\ep'}$ is invariant under $\cE\mapsto N^2-\cE$.
As an example, if $m=n=2$ this property can be readily verified using the explicit formula Eq.~\eqref{cZ22}.

As mentioned above, a duality relation of the form~\eqref{dual} clearly holds for an arbitrary chain
with global su$(m|n)$ symmetry, with Hamiltonian
\[
\hat\cH^{(m|n)}_{\ep\ep'}=\sum_{i\neq j}\big[c_{ij}\big(1-S_{ij}^{(m|n)}\big)+\tc_{ij}\big(1-\tS_{ij}^{(m|n)}\big)\big]
+\sum_i c_i\big(1-S_i^{\ep\ep'}\big)\,,
\]
where $c_{ij}$, $\tc_{ij}$ and $c_i$ are real constants. More precisely, from Eqs.~\eqref{SijcU}
and~\eqref{SicU} it now follows that
\[
\hat\cH^{(m|n)}_{-\ep,-\ep'}+{\cU^{(m|n)}}^{-1}\hat\cH^{(n|m)}_{\ep'\ep}\,\cU^{(m|n)}
=2\Big(\sum_{i\neq j}\big(c_{ij}+\tc_{ij}\big)+\sum_i c_i\Big)\equiv C\,,
\]
so that Eq.~\eqref{dual} becomes
\[
\hat\cZ^{(n|m)}_{\ep'\ep}(q)=q^{C}\,\hat\cZ^{(m|n)}_{-\ep,-\ep'}(q^{-1})\,.
\]

\section{Level density and spacings distribution}\label{sec:lev}

Having derived several explicit formulas for the partition function of the spin chain~\eqref{cH},
it is natural to inquire on the global properties of its spectrum. In fact, the partition function (and hence the spectrum) can be computed in a very efficient way for relatively large values of $N$ by expanding in
powers of $q$ Eqs.~\eqref{releven},
\eqref{cZoeA}, \eqref{cZeoA}, \eqref{cZooA} and~\eqref{cZ11++}--\eqref{cZ11-+}, together with
Eq.~\eqref{cZAkl}, with the help of a symbolic computation package. For instance,
using \mbox{\textsc{Mathematica}\texttrademark{}} on a personal computer it takes about five seconds to evaluate
the partition function $\cZ^{(2\mid 1)}_{--}$ for $N=40$ particles.

In the first place, our calculations for a wide range of values of $N$, $m$ and $n$
show that the energy levels of the chain~\eqref{cH} are equidistant, as already observed by us
in the purely bosonic or fermionic case~\cite{BFGR08}. More precisely, we have found that the
distance $\de\cE$ between consecutive levels is equal to one, except for the su$(1\vert 1)$
chain with $\ep=\ep'$, for which $\de\cE=2$ by Eqs.~\eqref{cZ11++} and~\eqref{cZ11--}.

\begin{figure}[h]
\psfrag{E}[Bc][Bc][1][0]{\footnotesize\,$\cE$}
\includegraphics[width=6.5cm]{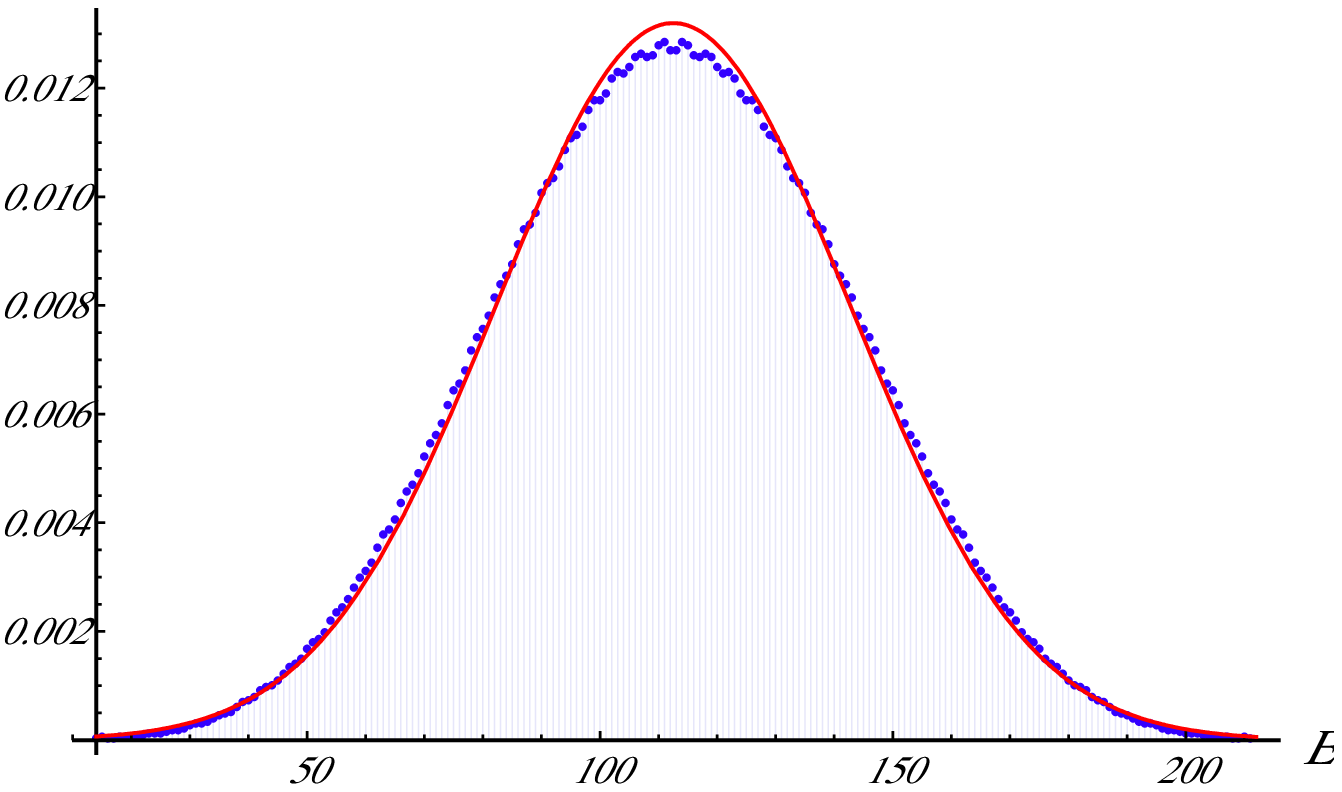}\hfill
\includegraphics[width=6.5cm]{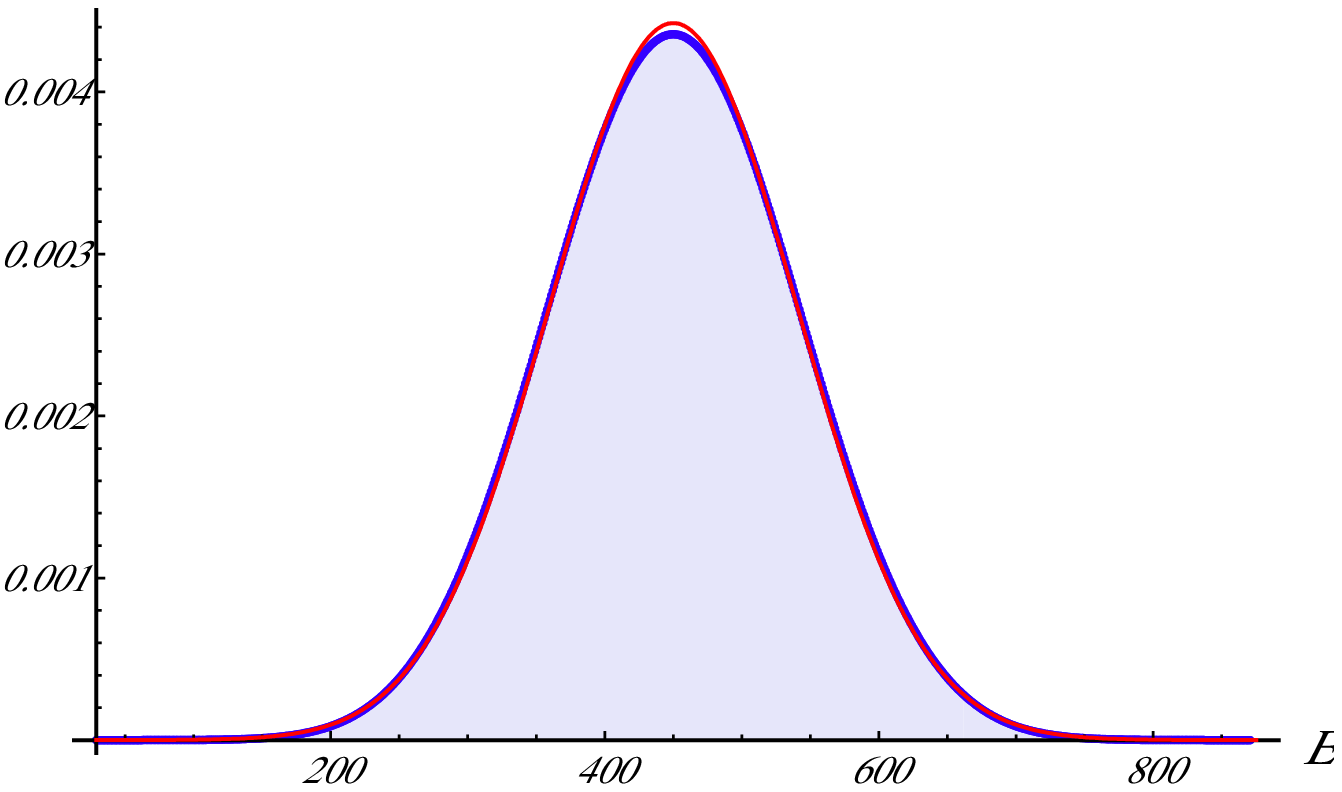}
\caption{Plot of the Gaussian distribution~\eqref{Gaussian} (continuous red line)
versus the LHS of Eq.~\eqref{fg} (blue dots) in the case $m=n=1$, $\ep'=-\ep=1$,
for $N=15$ (left) and  $N=30$ (right). The root mean square errors (normalized to the mean)
of the adjustments are $4.0\times 10^{-2}$ and $2.2\times 10^{-2}$, respectively.}
\label{fig:levdens}
\end{figure}

Secondly, our numerical calculations evidence that for sufficiently large $N$ ($N\gtrsim20$) the
normalized level density
\begin{equation}\label{f}
f(\cE)=(m+n)^{-N}\sum_{i=1}^L d_i\,\de(\cE-\cE_i)\,,
\end{equation}
where $\cE_1<\cdots<\cE_L$ are the distinct energy levels and $d_i$ is the degeneracy of $\cE_i$,
can be approximated with great accuracy by the Gaussian law
\begin{equation}\label{Gaussian}
g(\cE)=\frac{1}{\sqrt{2\pi}\si}\,\e^{-\frac{(\cE-\mu)^2}{2\si^2}}
\end{equation}
with parameters $\mu$ and $\si$ given by the mean and standard deviation of the chain's spectrum.
Since the energy levels are equidistant, this means that
\begin{equation}\label{fg}
\frac{d_i}{\de\cE(m+n)^N}\simeq g(\cE_i)\qquad (N\gg 1)\,.
\end{equation}
As an illustration, in Fig.~\ref{fig:levdens} we have plotted both sides of the latter equation
in the case $m=n=1$ and $\ep'=-\ep=1$ for $N=15$ and $N=30$. The approximate Gaussian character
of the level density is thus a property shared by all spin chains of HS type studied so far, both of type A and B~\cite{FG05,BB06,EFGR05,BFGR08,BFGR08b}.

Since, by Eq.~\eqref{fg}, the level density is asymptotically given by the Gaussian law~\eqref{Gaussian},
it is of interest to compute the mean energy and its variance as functions of $N$, $m$, $n$, $\ep$ and $\ep'$,
as has been done for other spin chains of HS type (including the purely fermionic version of the chain~\eqref{cH}).
Indeed, in Appendix~\ref{sec:meva} we show that $\mu$ and $\si^2$ are given by
\begin{align}
\mu&=\frac12\,\Big(1-\frac{m-n}{(m+n)^2}\Big)N(N-1)
+\Big(1-\frac{\ep\ms p_m+\ep'p_n}{m+n}\Big)\frac N2\,,\label{mu}\\[1mm]
\si^2&=\bigg(1-\frac{(m-n)^2}{(m+n)^4}\bigg)\frac N{36}(4N^2+6N-1)
+\frac{32\ms mn}{9(m+n)^4}\,N(N-1)(N-2)\notag\\
&\quad\;+\frac{2}{(m+n)^3}\,(n\ep\ms p_m-m\ep' p_n)N(N-1)\notag\\
&\quad\;+\frac{N}{4(m+n)^2}\bigg(\Big(\frac{m-n}{m+n}\Big)^2-(\ep\ms p_m+\ep'p_n)^2\bigg)\,,\label{si2final}
\end{align}
where $p_m,p_n\in\{0,1\}$ are the parities of $m$ and $n$, respectively.
In particular, when $N$ tends to infinity $\mu$ and $\si^2$ grow as $N^2$ and $N^3$, as
for the ordinary (non-supersymmetric) PF chains of type A and B (cf.~Refs.~\cite{BFGR08b,BFGR08}).

Another property of the chain's spectrum worth studying is the distribution of spacings between
consecutive ``unfolded'' levels. In general~\cite{GMW98}, the unfolding of the levels $\cE_i$ of a spectrum
is the mapping $\cE_i\mapsto\eta_i\equiv\eta(\cE_i)$, where $\eta(\cE)$ is the continuous
part of the cumulative level density
\[
F(\cE)\equiv\int_{-\infty}^\cE f(\cE')\d\cE'=\frac1{(m+n)^N}\,\sum_{i;\ms\cE_i\le\cE}d_i\,.
\]
It can be easily shown that the unfolded spectrum $\{\eta_i\}_{i=1}^L$ is uniformly distributed regardless of
the initial level density, making it meaningful to compare different spectra. In our case,
by the previous discussion we can take $\eta(\cE)$ as the cumulative Gaussian density~\eqref{Gaussian}, namely
\begin{equation}
\eta(\cE)=\int_{-\infty}^\cE g(\cE')\d\cE'=\frac12\,
\Big[
1+\erf\Big(\frac{\cE-\mu}{\sqrt 2\si}\Big)
\Big]\,.
\end{equation}
One then defines the normalized spacings
\[
s_i=(\eta_{i+1}-\eta_i)/\De\,,\qquad i=1,\dots,L-1\,,
\]
where $\De\equiv(\eta_{L}-\eta_1)/(L-1)$ is the mean spacing of the unfolded energies,
so that $\{s_i\}_{i=1}^{L-1}$ has unit mean.

As mentioned in the Introduction, the main motivation for studying the spacings density $p(s)$
lies in the Berry--Tabor conjecture, which states that the
distribution of spacings of a ``generic'' quantum integrable system
should obey Poisson's law $p(s)=\e^{-s}$. By contrast, for a chaotic
system the spacings distribution is expected to follow Wigner's surmise $p(s)=(\pi s/2)\ms\exp(-\pi s^2/4)$,
as is approximately the case for the Gaussian ensembles in random matrix theory~\cite{GMW98}.
A detailed study of the spacings distribution has been performed for many spin chains of HS type,
namely the PF chains of types A and B~\cite{BFGR08b,BFGR08},
the original (type A) HS chain~\cite{FG05,BFGR08b}, as well as the
supersymmetric version of the latter chain~\cite{BB06}.
The spacings distributions of all these chains, which turn out to be qualitatively very similar,
differ essentially from both Wigner's and Poisson's distributions. More precisely,
for the ordinary (non-supersymmetric) chains mentioned above we showed in our recent papers~\cite{BFGR08,BFGR08b}
that the cumulative spacings distribution $P(s)\equiv\int_0^s p(s')\d s'$ is
approximately given by
\begin{equation}\label{P}
P(s)\simeq 1-\frac{2}{\sqrt\pi\,\smax}\,\sqrt{\log\Big(\frac{\smax}s\Big)}\,,
\end{equation}
where $\smax$ is the maximum spacing. As a matter of fact, in Ref.~\cite{BFGR08} we proved
that the previous approximation holds for \emph{any} spectrum
$\cE_{\mathrm{min}}\equiv\cE_1<\cdots<\cE_L\equiv\cE_{\mathrm{max}}$,
provided that the following conditions are satisfied:

{\leftskip.75cm\parindent=0pt\setcounter{ex}{0}\parskip=6pt%
\cond The energies are equispaced, \emph{i.e.}, $\cE_{i+1}-\cE_i=\de\cE$ for $i=1,\dots,L-1$.

\cond The level density (normalized to unity) is approximately given by the Gaussian law~\eqref{Gaussian}.

\cond $\cE_{\mathrm{max}}-\mu\,,\,\mu-\cE_{\mathrm{min}}\gg\si$.

\cond $\cE_{\mathrm{min}}$ and $\cE_{\mathrm{max}}$ are approximately symmetric with respect to $\mu$, namely
$\vert\cE_{\mathrm{min}}+\cE_{\mathrm{max}}-2\mu\vert\ll\cE_{\mathrm{max}}-\cE_{\mathrm{min}}$.

}
Moreover, when these conditions are satisfied the maximum spacing can be estimated with great accuracy as
\begin{equation}\label{smax}
\smax=\frac{\cE_{\mathrm{max}}-\cE_{\mathrm{min}}}{\sqrt{2\pi}\,\si}\,.
\end{equation}
It should also be noted that Eq.~\eqref{P} is valid only for spacings $s\ge s_0$, where $s_0$ is the
unique zero of the RHS of this equation, namely
\begin{equation}\label{s0}
s_0=\smax\e^{-\frac\pi4\,\smax^2}\,.
\end{equation}
Since, by Eq.~\eqref{smax} and condition (iii), we have (for instance)
\[
\frac\pi4\,\smax^2>\frac{(\cE_{\mathrm{max}}-\mu)^2}{8\si^2}\gg 1\,,
\]
it follows that $s_0\ll\smax$. The validity of the above conditions for the chain~\eqref{cH} when $N\gg1$
was checked in Ref.~\cite{BFGR08} in the purely fermionic case,
which automatically implies their fulfillment in the purely bosonic case
on account of the duality relation~\eqref{dual}. For this reason, we shall assume in the rest
of this section that $m,n\ge 1$.  We shall see below that in this case conditions (i)--(iii)
are also satisfied when $N\gg1$, while condition (iv) only holds for
$m=n$. It turns out, however, that when one drops condition (iv)
the approximate formula~\eqref{P} still holds, albeit in a slightly smaller range.

\begin{figure}[h]
\centering
\psfrag{E}[Bc][Bc][1][0]{\footnotesize\,$\cE_i$}
\psfrag{s}[Bc][Bc][1][0]{\footnotesize\,$s_i$}
\psfrag{a}[Bl][Bl][1][0]{\footnotesize\,$\cE_{\mathrm{min}}$}
\psfrag{b}[Bl][Bl][1][0]{\footnotesize\,$\cE_{-}$}
\psfrag{c}[Bl][Bl][1][0]{\footnotesize\,$\cE_{+}$}
\psfrag{d}[Bl][Bl][1][0]{\footnotesize\,$\cE_{\mathrm{max}}$}
\psfrag{1}[Bc][Bc][1][0]{\footnotesize\,$s_{\mathrm{min}}$}
\psfrag{2}[Bc][Bc][1][0]{\footnotesize\,$s$}
\includegraphics[width=10cm]{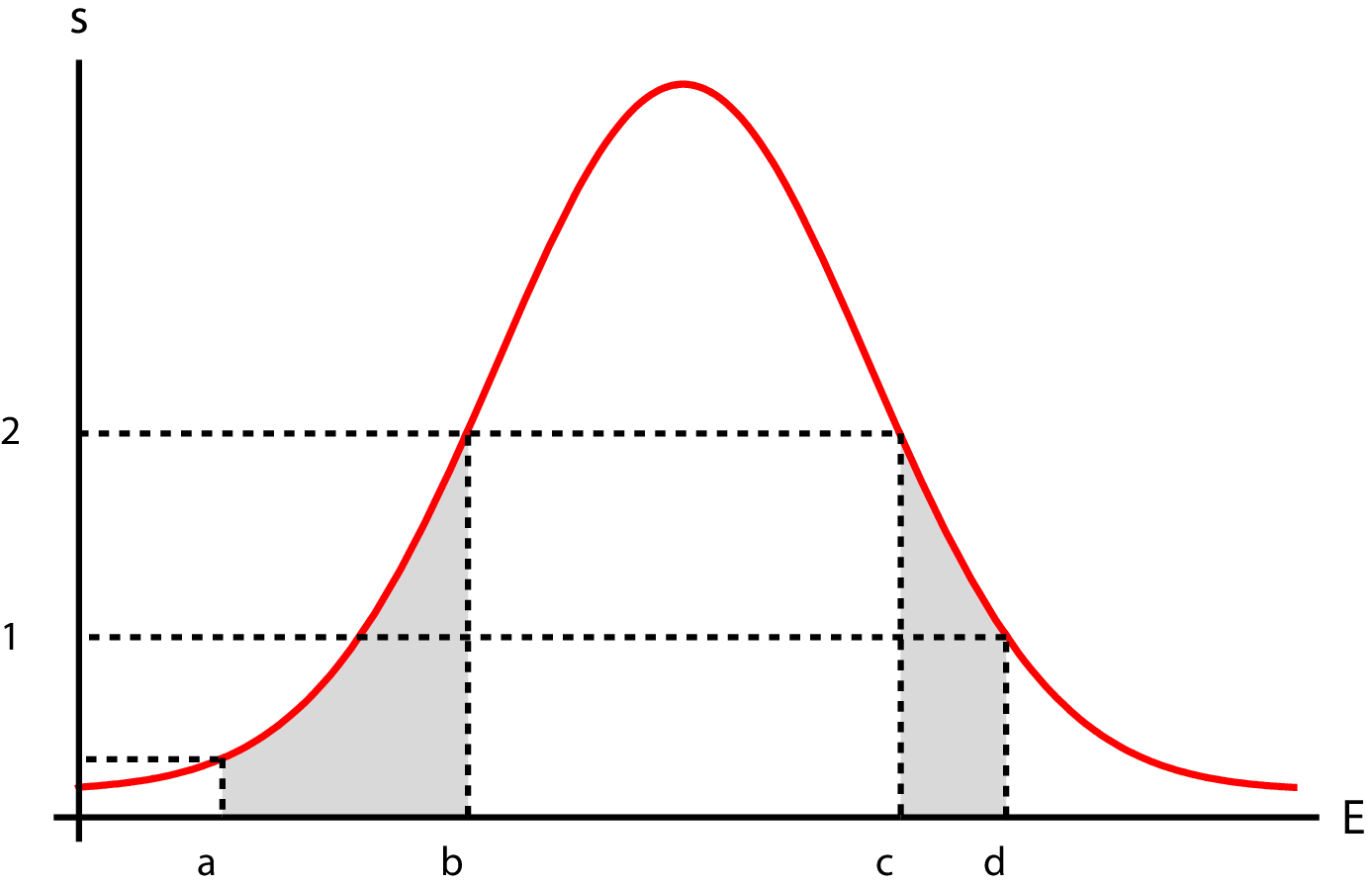}
\caption{Plot of the RHS of Eq.~\eqref{sicEi} showing the points $\cE_{\mathrm{min}}$, $\cE_{\mathrm{max}}$,
and the roots $\cE_{\pm}$ corresponding to a spacing $s\ge\smin$.}
\label{fig:plot-sE}
\end{figure}

Indeed, it was proved in Ref.~\cite{BFGR08} that conditions (i)--(iii) imply that the spacings $s_i$ are
approximately related to the energies $\cE_i$ by
\begin{equation}\label{sicEi}
s_i\simeq\smax\e^{-\frac{(\cE_i-\mu)^2}{2\si^2}}\,.
\end{equation}
Hence
\begin{equation}\label{Pdef}
P(s)\equiv\frac{\#(s_i\le s)}{L-1}=\frac1{L-1}\,
\Big[\#\big(\cE_{\mathrm{min}}\le\cE_i\le\cE_-\big)+\#\big(\cE_+\le\cE_i<\cE_{\mathrm{max}}\big)\Big]\,,
\end{equation}
where
\begin{equation}\label{cEpm}
\cE_\pm=\mu\pm\sqrt2\,\si\sqrt{\log\Big(\frac{\smax}s\Big)}
\end{equation}
are the roots of the equation $s=\smax\e^{-\frac{(\cE-\mu)^2}{2\si^2}}$, cf.~Fig.~\ref{fig:plot-sE}.
Using condition (i) to estimate the RHS of the latter equation we easily obtain
\begin{equation} \label{Pmax}
P(s)\simeq\frac1{(L-1)\ms\de\cE}\,\big[\max(\cE_--\cE_{\mathrm{min}},0)+\max(\cE_{\mathrm{max}}-\cE_+,0)\big]\,.
\end{equation}
If
\begin{multline}\label{smin}
\smin=\max\Big(\smax\e^{-\frac{(\cE_{\mathrm{min}}-\mu)^2}{2\si^2}},
\smax\e^{-\frac{(\cE_{\mathrm{max}}-\mu)^2}{2\si^2}}\Big)\\
\equiv\smax\e^{-\frac{1}{2\si^2}\left[\min(\mu-\cE_{\mathrm{min}},\cE_{\mathrm{max}}-\mu)\right]^2}\,,
\end{multline}
then for $s\ge\smin$ we have $\cE_-\ge\cE_{\mathrm{min}}$ and $\cE_+\le\cE_{\mathrm{max}}$; see
again Fig.~\ref{fig:plot-sE}. Hence, for $s\ge\smin$ Eq.~\eqref{Pmax} becomes
\begin{equation}\label{Pest}
P(s)\simeq\frac1{(L-1)\ms\de\cE}\,\big(\cE_--\cE_{\mathrm{min}}+\cE_{\mathrm{max}}-\cE_+\big)\,.
\end{equation}
Since (again by condition (i)) $\cE_{\mathrm{max}}-\cE_{\mathrm{min}}=(L-1)\ms\de\cE$, substituting
Eq.~\eqref{cEpm} into~\eqref{Pest} and using~\eqref{smax} we immediately arrive at the approximation~\eqref{P}
for~$P(s)$. We have thus shown that conditions (i)--(iii) imply that Eq.~\eqref{P} holds for $s\ge\smin$.
Note, finally, that the minimum spacing $\smin$ satisfies the inequalities
\[
s_0\le\smin\ll\smax\,.
\]
Indeed, the second inequality is an immediate consequence of Eq.~\eqref{smin} and condition~(iii).
As to the first one, by Eqs.~\eqref{smax}, \eqref{s0} and \eqref{smin} we have
\begin{multline*}
\log\Big(\frac{s_0}{\smax}\Big)=-\frac1{8\si^2}\,(\cE_{\mathrm{max}}-\cE_{\mathrm{min}})^2\\
\le-\frac1{8\si^2}\,\big[2\min(\mu-\cE_{\mathrm{min}},\cE_{\mathrm{max}}-\mu)\big]^2
=\log\Big(\frac{\smin}{\smax}\Big)\,,
\end{multline*}
where the equality holds if and only if $\cE_{\mathrm{min}}$ and $\cE_{\mathrm{max}}$ are symmetric about $\mu$.

Let us next verify that the spectrum of the chain~\eqref{cH} satisfies conditions (i)--(iii) when $N\gg 1$
and $m,n\ge 1$. We have already seen at the beginning of this section that conditions (i) and (ii) hold
for sufficiently large $N$. In order to check the validity of condition (iii), we need to
compute in closed form the chain's maximum and minimum energies.
By equation~\eqref{EEE}, the minimum energy is given by
\[
\cE_{\mathrm{min}}=\lim_{a\to\infty}\frac1a\,(E_{\mathrm{min}}-E^{\mathrm{sc}}_{\mathrm{min}})\,,
\]
where $E^{\mathrm{sc}}_{\mathrm{min}}$ and $E_{\mathrm{min}}$ are the minimum energies of the
scalar and spin dynamical models~\eqref{Hsc} and~\eqref{H}, respectively. {}From the discussion
in Section~\ref{sec:spectrum} it follows that $E^{\mathrm{sc}}_{\mathrm{min}}=E_0$, so that
(cf.~Eq.~\eqref{Eks}) $\cE_{\mathrm{min}}$ is the minimum value of $\vert\bk\vert$, where $\bk$ is any multi-index
compatible with conditions (i)--(iii) in Section~\ref{sec:spectrum}. If $m>1$ we can obviously take
$\bk=0$ (and, for instance, $\bs=\big((s,0),\dots,(s,0)\big)$, where $s$ is any positive spin component),
so that $\cE_{\mathrm{min}}=0$ in this case. Similarly, when $m=\ep=1$ the minimum energy is again zero,
since the pair $\bk=0$ and $\bs=\big((0,0),\dots,(0,0)\big)$ satisfies the required conditions. However,
if $m=1$ and $\ep=-1$, condition (iii) implies that $k_i$ must be odd when $s_i^1=0$. Hence,
the multi-index $\bk$ yielding the minimum energy is of the form
\[
\bk=\begin{cases}
(\overbrace{\vrule height10pt width0pt 1,\dots,1}^{N-\overline n(0)}
\,,\overbrace{\vrule height10pt width0pt0,\dots,0}^{\overline n(0)})\,,&\qquad\overline n(0)<N\,,\\[2mm]
(0,\dots,0)\,,&\qquad\overline n(0)\ge N\,,
\end{cases}
\]
with $\cE_{\mathrm{min}}=\max\big(N-\overline n(0),0\big)$, where the notation $\overline n(k_j)$
is defined by the fermionic analogue of Eq.~\eqref{mbar}.
(A compatible vector $\bs$ is obtained by filling the rightmost $\overline n(0)$ components
with all possible nonnegative fermionic spin values, and the rest (if any) with the unique bosonic
spin $s_i^1=0$.) In summary, if $m,n\ge 1$ the minimum energy is given by
\begin{equation}\label{cEmin}
\cE_{\mathrm{min}}=
\begin{cases}
0\,,&\qquad m>1,\text{ or } m=\ep=1\,,\\[1mm]
\max\big(N-\overline n(0),0\big)\,,&\qquad m=-\ep=1\,.
\end{cases}
\end{equation}
As to the maximum energy, the duality relation~\eqref{dual} and the previous equation imply that
\begin{equation}\label{cEmax}
\cE_{\mathrm{max}}=
\begin{cases}
N^2\,,&\qquad n>1,\text{ or } n=-\ep'=1\,,\\[1mm]
N^2-\max\big(N-\overline m(0),0\big)\,,&\qquad n=\ep'=1\,.
\end{cases}
\end{equation}
{}From Eqs.~\eqref{mu}, \eqref{si2final} and the last two equations, it easily follows that
both $(\cE_{\mathrm{max}}-\mu)/\si$ and $(\mu-\cE_{\mathrm{min}})/\si$ are $O(N^\frac12)$ as $N\to\infty$,
so that condition (iii) above is indeed satisfied.
On the other hand, by Eqs.~\eqref{mu}, \eqref{cEmin} and~\eqref{cEmax} we have
\[
\frac{\vert\cE_{\mathrm{min}}+\cE_{\mathrm{max}}-2\mu\vert}{\cE_{\mathrm{max}}-\cE_{\mathrm{min}}}=
\frac{\vert m-n\vert}{(m+n)^2}+O(N^{-1})\qquad\qquad (m,n\ge 1)\,,
\]
so that condition (iv) is not satisfied unless $m=n$, as claimed above.

\begin{figure}[h]
\centering
\psfrag{P}[Bc][Tc][1][0]{\footnotesize $P(s)$}
\psfrag{s}[Bc][Bc][1][0]{\footnotesize $s$}
\includegraphics[width=6.5cm]{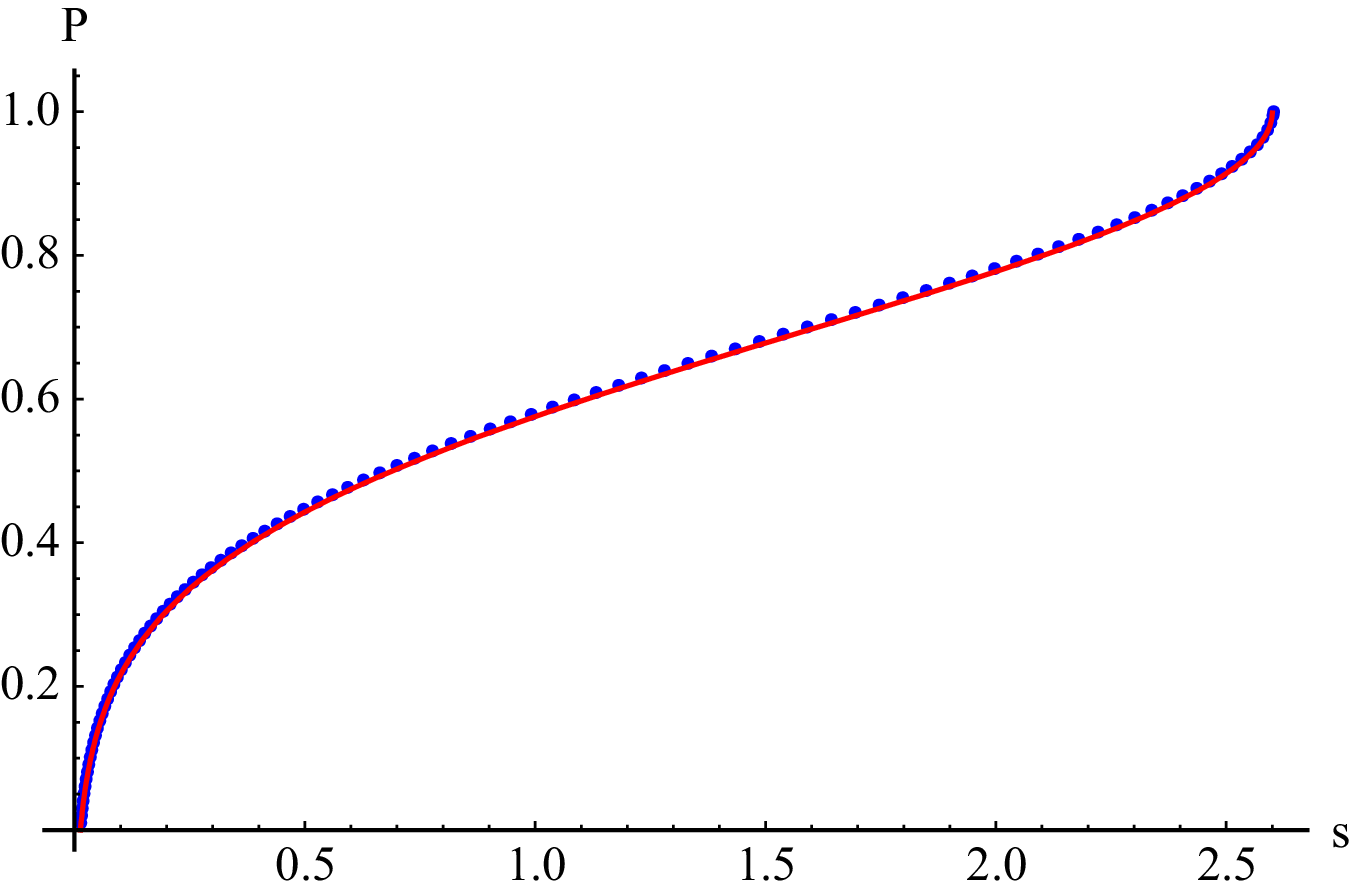}\hfill
\includegraphics[width=6.5cm]{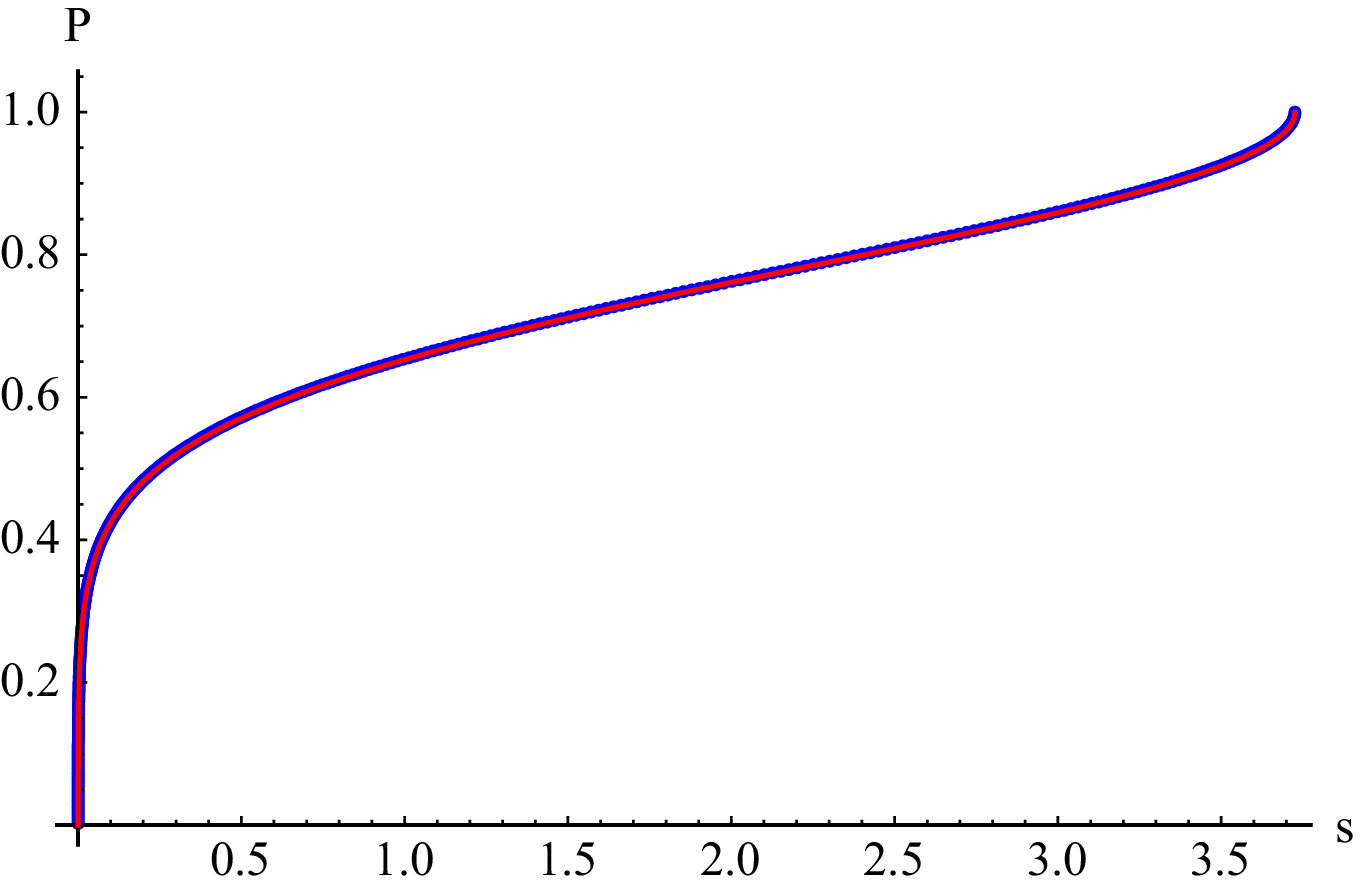}\\[2mm]
\includegraphics[width=6.5cm]{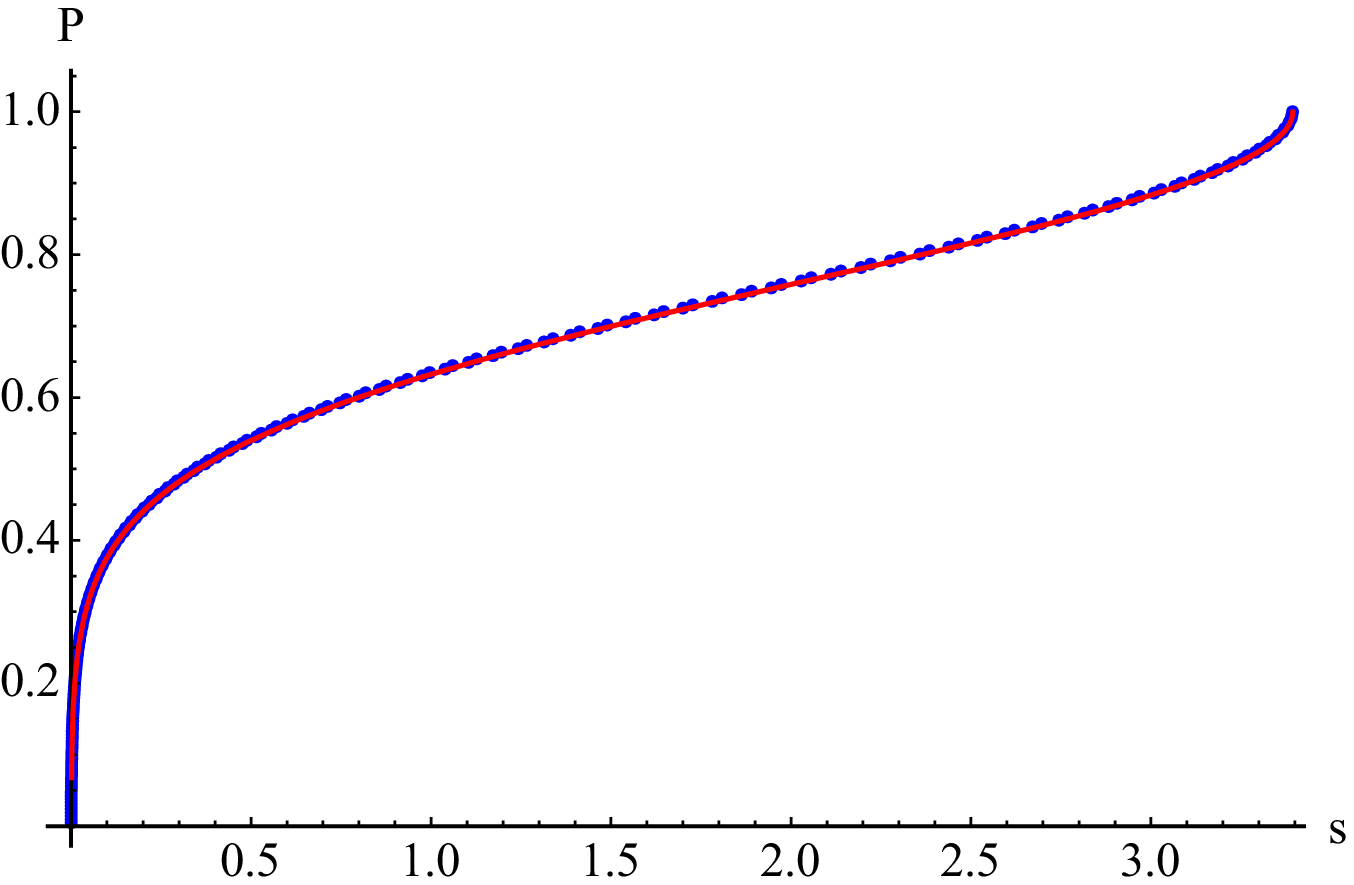}\hfill
\includegraphics[width=6.5cm]{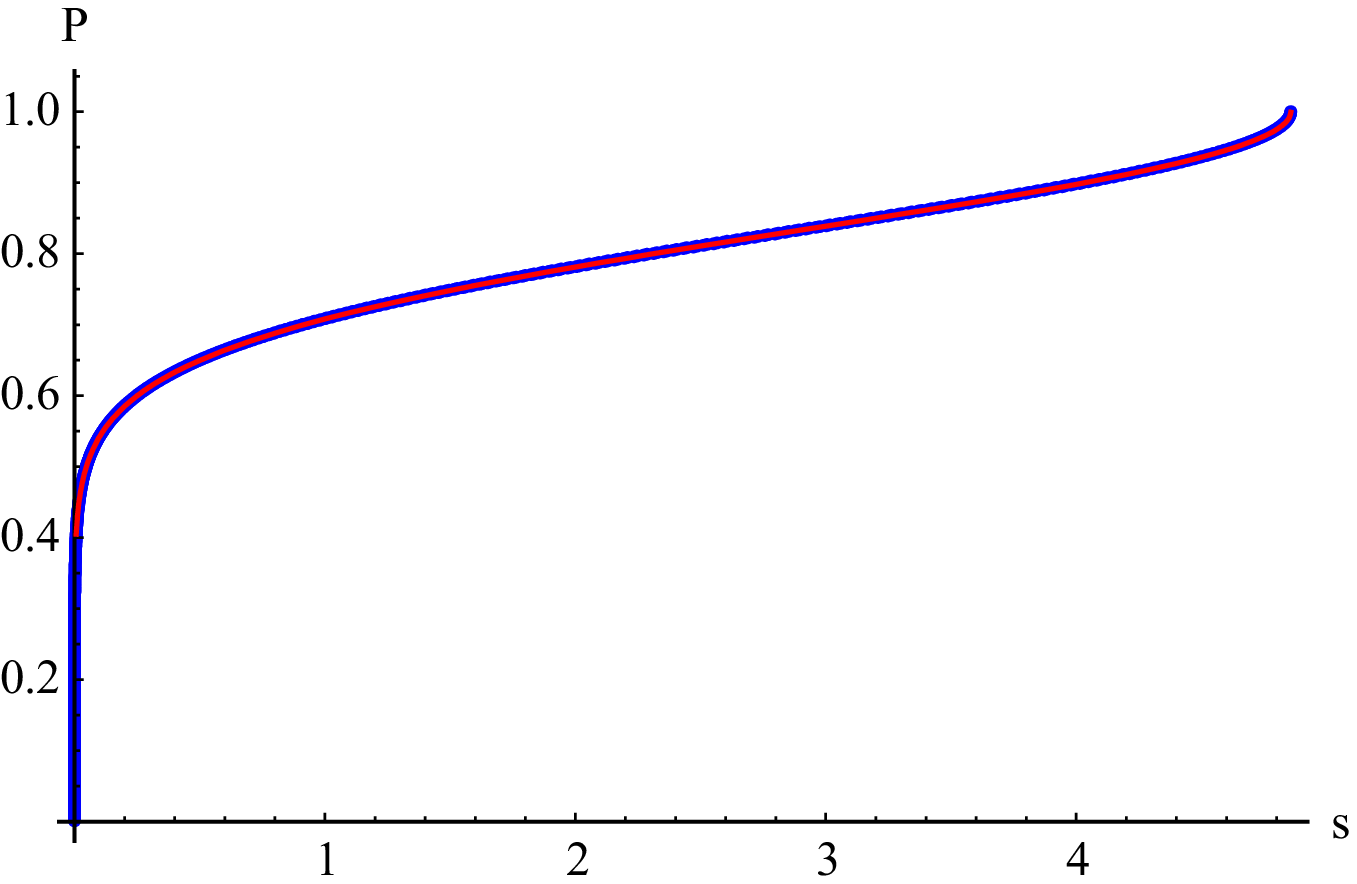}
\caption{Cumulative spacings distribution $P(s)$ (blue dots) and its analytic
approximation~\eqref{P}-\eqref{smax} (continuous red line) in the cases $m=n=-\ep=\ep'=1$ (top) and $m=2$, $n=\ep=\ep'=1$
(bottom) for $N=15$ (left) and $N=30$ (right).}
\label{fig:sp}
\end{figure}

We have verified that the approximation~\eqref{P}-\eqref{smax} is in excellent agreement with the numerical data obtained
using our exact formulas for the partition function for many different values of $m,n\ge 1$ and $N\ge15$.
For instance, in Fig.~\ref{fig:sp} we plot the cumulative spacings distribution $P(s)$
versus its analytic approximation~\eqref{P} in the cases $m=n=-\ep=\ep'=1$ and $m=2$, $n=\ep=\ep'=1$
for $N=15$ and $N=30$ spins. The respective root mean square errors (normalized to the mean)
drop from $9.4\times 10^{-3}$ and $4.6\times 10^{-3}$ for $N=15$, to $2.4\times 10^{-3}$
and $1.1\times 10^{-3}$ for $N=30$. It is worth noting that the approximation~\eqref{P}-\eqref{smax} for the
cumulative spacings distribution contains no adjustable parameters, since the maximum
spacing $\smax$ is completely determined by Eqs.~\eqref{si2final}, \eqref{cEmin} and~\eqref{cEmax}.
In fact, from the previous equations one easily obtains the following asymptotic formula
for the maximum spacing in the genuinely supersymmetric case $m,n\ge 1$:
\[
\smax=\frac3{\sqrt{2\pi}}\,\Big(1+\frac{34mn-m^2-n^2}{(m+n)^4}\Big)^{\!\!-\frac12}N^{\frac12}+O(N^{-\frac12})\,.
\]
Thus the behavior of $\smax$ for large $N$ is qualitatively the same as in the
non-supersymmetric cases studied in Ref.~\cite{BFGR08}.

\begin{ack}
This work was partially supported by the DGI under grant
no.~FIS2005-00752, and by the Complutense University of Madrid and the
DGUI under grant no.~GR74/07-910556. JCB acknowledges the financial support of the
Spanish Ministry of Science and Innovation through an FPU scholarship.
\end{ack}

\begin{appendix}
\section{Some useful $q$-number identities}\label{sec:q}

In this appendix we shall collect several identities involving $q$-numbers
that are used in the simplification of the partition function
$\cZ^{(m|n)}_{\ep\ep'}$ of the chain~\eqref{cH} when either $m$ or $n$ are
odd. Given a real number $q\in(0,1)$ and a nonnegative integer $k$, we define
the symbols $(q)_k$ and $\qsb kq$ by
\begin{equation}\label{qkkq}
(q)_k=\prod_{i=1}^k(1-q^i)\,,\qquad\qsb kq=\frac{1-q^k}{1-q}=\sum_{i=0}^{k-1}q^i\,,
\end{equation}
with $(q)_0\equiv 1$. The $q$-\emph{factorial} of $k$ is then defined as
\[
\qsb kq!=\prod_{i=1}^k\qsb iq\,.
\]
Finally, if $k,l$ are nonnegative integers with $k\ge l$, we define the $q$-\emph{binomial coefficient}
\[
\qbinom klq=\frac{\qsb kq!}{\qsb lq!\,\qsb{k-l}q!}\,.
\]
{}From the above definitions it immediately follows the useful equality
\begin{equation}\label{qbinomid}
\qbinom klq=\frac{(q)_k}{(q)_l(q)_{k-l}}\,.
\end{equation}
An important identity satisfied by the $q$-binomial coefficients is the
following instance of Newton's $q$-binomial formula~\cite[Eq.~(8)]{Ci79}:
\begin{equation}
  \label{qbinom}
  \sum_{i=0}^k\qbinom kiq q^{\binom i2}\,x^i=\prod_{i=0}^{k-1}(1+x\ms q^i)\,.
\end{equation}
From the previous equation with $q$ replaced by $q^2$ and $x$ by $q$ we obtain
the formula
\begin{equation}
  \label{qbinom2}
\sum_{i=0}^k\qbinom ki{q^2}\,q^{i^2}=\prod_{i=0}^{k-1}(1+q^{2i+1})\,.
\end{equation}
Replacing $q$ by $q^2$ and setting $x=q^{-1}$ in Eq.~\eqref{qbinom}
we obtain the analogous relation
\begin{equation}
  \label{qbinom3}
\sum_{i=0}^k\qbinom ki{q^2}\,q^{i(i-2)}=\prod_{i=0}^{k-1}(1+q^{2i-1})\,.
\end{equation}
The last three identities are used in Section \ref{sec:rel} to compute the
partition function~$\cZ^{(1|1)}_{\ep\ep'}$.

Likewise, if $\de_q$ is the $q$-dilation operator,
whose action on a smooth function $f:\RR\to\RR$ is given by $(\de_qf)(x)=f(qx)$,
we have~\cite[Eq.~(11)]{Ci79}
\[
\sum_{i=0}^k\qbinom kiq x^i=(x+\de_q)^k\cdot 1\,,
\]
where $1$ denotes the constant function identically equal to $1$. Equating the
coefficient of $x^i$ in both sides of the previous equation we obtain the
remarkable identity
\begin{equation}
  \label{qbinompoly}
  \qbinom kiq = q^{-\binom i2}\,\tau_i(1,q,\dots,q^{k-1})\,,
\end{equation}
where $\tau_i(x_1,\dots,x_k)$ denotes the elementary symmetric polynomial of degree $i$ in
$k$ variables, defined by
\begin{equation}
  \label{tau}
  \tau_i(x_1,\dots,x_k)=\sum_{1\le j_1<\cdots<j_i\le k}x_{j_1}\cdots x_{j_i}\,.
\end{equation}
In particular, from Eq.~\eqref{qbinompoly} it easily follows that the
$q$-binomial coefficient $\qbinom kiq$ is a polynomial of degree $i(k-i)$ in $q$.

\section{Computation of the mean and variance of the energy}\label{sec:meva}

In this appendix we shall compute the mean and variance of the energy of the chain~\eqref{cH}
for arbitrary values of $N$, $m$, $n$, $\ep$ and $\ep'$. In the purely  fermionic
case, these quantities were computed in Ref.~\cite{BFGR08}
using the formulas for the traces (of products) of the operators $S_{ij}$ and $S_i$
given in Ref.~\cite{EFGR05}. In the genuinely supersymmetric case, one should evaluate the traces
of products involving the more general operators $\Sij$ and $\Si$. Indeed,
the traces of $\Sij$ and $\Sij S_{kl}^{(m|n)}$, which are the only ones appearing in
the chain of type A, were computed in detail in Ref.~\cite{BB06}.
The calculation of the traces involving the spin reversal operator $\Si$ is totally analogous,
so that we shall limit ourselves to listing the final results in Table~\ref{table:traces}.
In this table $p_m,p_n\in\{0,1\}$ are the parities of $m$ and $n$, respectively, and we have omitted
the traces of products which reduce to the identity matrix or to one of the entries.

\smallskip\begin{table}[h]
\caption{Traces of products of the  spin operators $\Sij$, $\tSij$ and $\Si$.}\label{table:traces}
\begin{tabular}{lll}\hline
  \vrule height 15pt depth 9pt width0pt Operator & Trace\\ \hline
$\Si\ms$ & $(m+n)^{N-1}(\ep p_m+\ep' p_n)$\\ \hline
$\Sij\ms$,\, $\tSij$ & $(m-n)(m+n)^{N-2}$\\ \hline
$\Si\Sj$ & $(m+n)^{N-2}(\ep p_m+\ep' p_n)^2$\\ \hline
$\Sij\Sk\ms$,\, $\tSij\Sk\ms$ & $(m-n)(m+n)^{N-3}(\ep p_m+\ep' p_n)$\\ \hline
$\Sij\Si\ms$,\, $\tSij\Si\ms$ & $(m+n)^{N-2}(\ep p_m-\ep' p_n)$\\ \hline
$\Sij\Skl\ms$,\, $\Sij\tSkl\ms$,\, $\tSij\tSkl$ & $(m-n)^2(m+n)^{N-4}$\\ \hline
$\Sij\Sil\ms$,\, $\Sij\tSil\ms$,\, $\tSij\tSil$\hspace{2em} & $(m+n)^{N-2}$\\ \hline
\end{tabular}
\end{table}

In the first place, calling
\[
h_{ij}=(\xi_i-\xi_j)^{-2}\,,\quad \tilh_{ij}=(\xi_i+\xi_j)^{-2}\,,\quad
h_i=\be\,\xi_i^{-2}\,,
\]
and using the formulas in Table~\ref{table:traces} for the traces of the spin operators
we easily obtain
\begin{align}
  \mu &= \frac{\tr\cH^{(m|n)}_{\ep\ep'}}{(m+n)^N}\notag\\
  \label{muexp}
  &= \Big(1-\frac{m-n}{(m+n)^2}\Big)\sum_{i\neq
    j}(h_{ij}+\tilh_{ij})+\Big(1-\frac{\ep\ms p_m+\ep'p_n}{m+n}\Big)\sum_i h_i\,.
\end{align}
The sums in the previous expression were evaluated in Ref.~\cite{BFGR08}, with the result
\[
\sum_{i\neq j}(h_{ij}+\tilh_{ij})=\frac12\,N(N-1)\,,\qquad
\sum_i h_i=\frac N2\,.
\]
Inserting these values into~\eqref{muexp} we immediately arrive at Eq.~\eqref{mu} for the mean
energy $\mu$.

The evaluation of the standard deviation of the energy
\begin{equation}\label{si2}
\si^2=\frac{\tr\left[\big(\cH^{(m|n)}_{\ep\ep'}\big)^2\right]}{(m+n)^N}-\mu^2\,,
\end{equation}
which is considerably more involved, is also performed in two steps. We begin by expanding Eq.~\eqref{si2}
and simplifying the resulting expression using Eq.~\eqref{muexp} and the trace formulas in Table~\ref{table:traces}.
After a long but straightforward calculation we obtain
\begin{align}
\si^2&=\bigg(1-\frac{(m-n)^2}{(m+n)^4}\bigg)\Big(2\sum_{i\neq j}(h_{ij}^2+\tilh_{ij}^2)
+\sum_ih_i^2\Big)\notag\\
&\quad+\frac4{(m+n)^2}\bigg(\Big(\frac{m-n}{m+n}\Big)^2-(\ep\ms p_m+\ep'p_n)^2\bigg)
\Big(\frac14\sum_ih_i^2-\sum_{i\neq j}h_{ij}\tilh_{ij}\Big)\notag\\
&\quad+\frac8{(m+n)^3}\,(n\ep\ms p_m-m\ep' p_n)\sum_{i\ne j}(h_{ij}+\tilh_{ij})h_i\notag\\
&\quad+\frac{16\ms mn}{(m+n)^4}{\sum_{i,j,k}}^{\ms\prime}(h_{ij}+\tilh_{ij})(h_{ik}+\tilh_{ik})\,,
\label{si2sum}
\end{align}
where the symbol ${\sum\limits_{i,j,k}}^{\!\prime}$ denotes summation over $i\ne j\ne k\ne i$. The sums in the
first two lines of Eq.~\eqref{si2sum}, which were evaluated in Ref.~\cite{BFGR08}, are given by
\begin{align}
2\sum_{i\neq j}(h_{ij}^2+\tilh_{ij}^2)+\sum_ih_i^2&=\frac N{36}\,(4N^2+6N-1)\,,\label{sum1}\\
\frac14\sum_ih_i^2-\sum_{i\neq j}h_{ij}\tilh_{ij}&=\frac N{16}\,.\label{sum2}
\end{align}
In order to evaluate the last two sums in~\eqref{si2sum}, we recall from
Appendix~A of Ref.~\cite{BFGR08} the following identities involving the zeros
$y_i=\xi_i^2/2$ of the Laguerre polynomial $L_N^{\be-1}$ and their differences $y_{ij}\equiv y_i-y_j$\,:
\begin{subequations}\label{eqs}\allowdisplaybreaks
\begin{align}
\sum_i\frac1{y_i^2}&=\frac{N(N+\be)}{\be^2(\be+1)}\label{eqa}\\[1mm]
\sum_i\frac1{y_i}&=\frac N\be\,,\label{eqb}\\[1mm]
\sum_iy_i&=N(N+\be-1)\,,\label{eqc}\\[1mm]
\sum_iy_i^2&=N(N+\be-1)(2N+\be-2)\,,\label{eqd}\\[1mm]
\sum_{j,j\neq i}\frac{y_i}{y_{ij}}&=\frac12\,(y_i-\be)\,,\label{eqe}\\[1mm]
\sum_{j,j\neq i}\frac{y_i^2}{y_{ij}^2}&=\frac1{12}\,\big(-y_i^2+2(2N+\be)y_i-\be(\be+4)\big)\,,\label{eqf}\\[1mm]
\sum_{i\neq j}\frac1{y_{ij}^2}&=\frac{N(N-1)}{4(\be+1)}\,,\label{eqg}\\[1mm]
\sum_{i\neq j}\frac{y_i}{y_{ij}^2}&=\frac14\,N(N-1)\,,\label{eqh}\\[1mm]
\sum_{i\neq j}\frac{y_i^2}{y_{ij}^4}&=\frac{N(N-1)\big((2N+5)\be+2N+14\big)}{144(\be+1)}\,.\label{eqi}
\end{align}
\end{subequations}
Multiplying Eq.~\eqref{eqe} by $y_i^{-2}$ and $y_i$, summing over $i$ the resulting identities, and
using~\eqref{eqa}--\eqref{eqd}, one easily obtains
\begin{equation}\label{ceq1}
\sum_{i\neq j}\frac{y_i^{-1}}{y_{ij}}=-\frac{N(N-1)}{2\be(\be+1)}\,,\qquad
\sum_{i\neq j}\frac{y_i^2}{y_{ij}}=N(N-1)(N+\be-1)\,.
\end{equation}
Similarly, summing Eq.~\eqref{eqf} over $i$ and using~\eqref{eqc}-\eqref{eqd} we get
\begin{equation}\label{ceq2}
\sum_{i\neq j}\frac{y_i^2}{y_{ij}^2}=\frac1{12}\,N(N-1)(2N+3\be+2)\,.
\end{equation}
The sum in the third line of Eq.~\eqref{si2sum} can now be easily computed. Indeed,
\begin{align}
\sum_{i\ne j}(h_{ij}+\tilh_{ij})h_i &=\frac\be2\,\sum_{i\ne j}\frac{y_i+y_j}{y_i\,y_{ij}^2}\notag\\
&=\be\sum_{i\ne j}\frac1{y_{ij}^2}-\frac\be2\,\sum_{i\neq j}\frac{y_i^{-1}}{y_{ij}}
=\frac14\,N(N-1)\,,\label{sum3}
\end{align}
where we have used Eqs.~\eqref{eqg} and~\eqref{ceq1}. Turning next to the last sum in Eq.~\eqref{si2sum},
we have
\begin{align}
{\sum_{i,j,k}}^{\ms\prime}(h_{ij}+\tilh_{ij})(h_{ik}+\tilh_{ik})&=
{\sum_{i,j,k}}^{\ms\prime}\,\frac{(y_i+y_j)(y_i+y_k)}{y_{ij}^2\,y_{ik}^2}\notag\\
&={\sum_{i,j,k}}^{\ms\prime}\,\frac{4y_i^2}{y_{ij}^2\,y_{ik}^2}
-{\sum_{i,j,k}}^{\ms\prime}\,\frac{4y_i}{y_{ij}^2\,y_{ik}}
+{\sum_{i,j,k}}^{\ms\prime}\,\frac1{y_{ij}\,y_{ik}}\,.\label{sum4}
\end{align}
The last sum in the previous equation clearly vanishes on account of the identity
\[
\frac1{(z_1-z_2)(z_1-z_3)}+\frac1{(z_2-z_3)(z_2-z_1)}+\frac1{(z_3-z_1)(z_3-z_2)}=0\,,
\]
valid for arbitrary $z_1\ne z_2\ne z_3\ne z_1$. On the other hand, the first two sums in
the RHS of Eq.~\eqref{sum4} can be computed using Eqs.~\eqref{eqs}--\eqref{ceq2}. Indeed,
\begin{align*}
{\sum_{i,j,k}}^{\ms\prime}\,\frac{4y_i^2}{y_{ij}^2\,y_{ik}^2}&=
\sum_{i\ne j}\frac4{y_{ij}^2}\bigg(\sum_{k,k\ne i}\frac{y_i^2}{y_{ik}^2}-\frac{y_i^2}{y_{ij}^2}\bigg)\\
&=\frac13\,\sum_{i\ne j}\frac1{y_{ij}^2}\big(-y_i^2+2(2N+\be)y_i-\be(\be+4)\big)-4\sum_{i\ne j}\frac{y_i^2}{y_{ij}^4}\\
&=\frac29\,N(N-1)(N-2)\,,
\end{align*}
where we have used Eqs.~\eqref{eqf}--\eqref{eqh} and~\eqref{ceq2}. Likewise,
\begin{align*}
{\sum_{i,j,k}}^{\ms\prime}\,\frac{y_i}{y_{ij}^2\,y_{ik}}&=
\sum_{i\ne j}\frac1{y_{ij}^2}\bigg(\sum_{k,k\ne i}\frac{y_i}{y_{ik}}-\frac{y_i}{y_{ij}}\bigg)
=\frac12\sum_{i\ne j}\frac{y_i-\be}{y_{ij}^2}-\sum_{i\ne j}\frac{y_i}{y_{ij}^3}\\[1mm]
&=\frac12\sum_{i\ne j}\frac{y_i-\be}{y_{ij}^2}-\frac12\sum_{i\ne j}\frac1{y_{ij}^2}
=\frac12\sum_{i\ne j}\frac{y_i}{y_{ij}^2}-\frac{\be+1}2\sum_{i\ne j}\frac1{y_{ij}^2}=0\,,
\end{align*}
on account of Eqs.~\eqref{eqg}-\eqref{eqh}. The LHS of Eq.~\eqref{sum4} thus reduces to
\begin{equation}\label{sum4final}
{\sum_{i,j,k}}^{\ms\prime}(h_{ij}+\tilh_{ij})(h_{ik}+\tilh_{ik})=\frac29\,N(N-1)(N-2)\,.
\end{equation}
Substituting Eqs.~\eqref{sum1}, \eqref{sum2}, \eqref{sum3} and \eqref{sum4} into~\eqref{si2sum}
we finally obtain the expression~\eqref{si2final} for $\si^2$ in Section~\ref{sec:lev}.
\end{appendix}


\end{document}